\documentclass[a4,letterpaper,11pt]{article}

\pdfoutput=1 

\usepackage{jheppub} 

\usepackage[T1]{fontenc}
\usepackage{subfig}
\usepackage{lmodern}
\usepackage{amsmath}
\usepackage{amssymb}
\usepackage{graphicx}
\usepackage{xspace}
\usepackage{slashed}
\usepackage{multirow}

\usepackage{epstopdf}

\newcommand{\nn}{\nonumber}

\newcommand{\lp}{\Big{(}}
\newcommand{\rp}{\Big{)}}
\newcommand{\lbc}{\Big{\lbrace}}
\newcommand{\rbc}{\Big{\rbrace}}
\newcommand{\ba}{\begin{align}}
\newcommand{\ea}{\end{align}}
\newcommand{\bea}{\begin{eqnarray}}
\newcommand{\eea}{\end{eqnarray}}

\allowdisplaybreaks[1]

\preprint{\begin{flushright}
MIT--CTP 5243
\end{flushright}}

\title{Effective Field Theory for Jet substructure in heavy ion collisions}

\author{Varun Vaidya}

\affiliation{Center for Theoretical Physics, Massachusetts Institute of Technology, Cambridge, MA~02139, U.S.A.}

\abstract{I develop an Effective Field Theory (EFT) framework to compute jet substructure observables for heavy ion collision experiments. 
As an illustration, I consider dijet events that accompany the formation of a weakly coupled Quark Gluon Plasma(QGP) medium in a heavy ion collision and look at an observable insensitive to jet selection bias: the simultaneous measurement of jet mass along with the transverse momentum imbalance between the jets that are groomed to remove soft radiation. Treating the jet as an open quantum system, I write down a factorization formula within the SCET(Soft Collinear Effective Theory) framework in the forward scattering regime. The physics of the medium is encoded in a universal soft field correlator while the jet-medium interaction is captured by a medium induced jet function. The factorization formula leads to a Lindblad type equation for the evolution of the reduced density matrix of the jet in the Markovian approximation. The solution for this equation allows a resummation of large logarithms that arise due to the final state measurements imposed while simultaneously summing over multiple incoherent interactions of the jet with the medium.
}

\begin{document}
\maketitle

\pagebreak

\section{Introduction}

The natural final state of high energy hadronic or nuclear collisions are sprays of collimated particles consisting of hadrons and/or electrons. They are formed in an initial hard scattering, by which we mean a collision with a large transfer of momentum, followed by subsequent parton evolution know as a shower and fragmentation. Due to the large energies involved in jet production, they can be studied via perturbative QCD since QCD is weakly coupled at high virtualities of a parton. Therefore the calculation of the initial production of jets is under perturbative control, which makes jets powerful tools to probe the properties of the quark-gluon plasma (QGP) in heavy ion collisions. 

This is based on the premise, which is now widely accepted that heavy ion collisions are the laboratory for the creation and study of the Quark Gluon Plasma medium. The high energy collision of nuclei both at RHIC and the LHC creates sufficiently energetic partons that can escape confinement from color neutral hadrons and give rise to a strongly/weakly coupled soup of quarks and gluons known as the Quark Gluon Plasma medium which, in thermal equilibrium is mainly characterized by its temperature. We can think of this plasma as consisting of soft partons with typical energy of the order of the temperature of the medium which is usually much lower than the center of mass energy of the initiating nuclear collision. These stopping collisions which create the QGP are accompanied by hard interactions which create highly energetic partons which eventually form jets. These jets then have to traverse through a region of the hot QGP as they evolve and hence they get modified in heavy ion collision, compared with proton-proton collisions, due to the jet-medium interaction. 

We would therefore like to study the modification of jet substructure for the same hard event in heavy ion collision(HIC) compared to a pp collision as a tool for extracting the properties of the medium. However, the selection of jets in a HIC suffers from the so called jet selection bias which is related to a phenomenon of jet quenching, that has been extensively studied in  literature\cite{Gyulassy:1993hr,Wang:1994fx,Baier:1994bd,Baier:1996kr,Baier:1996sk,Zakharov:1996fv,Zakharov:1997uu,Gyulassy:1999zd,Gyulassy:2000er,Wiedemann:2000za,Guo:2000nz,Wang:2001ifa,Arnold:2002ja,Arnold:2002zm,Salgado:2003gb,Armesto:2003jh,Majumder:2006wi,Majumder:2007zh,Neufeld:2008fi,Neufeld:2009ep} and entails the systematic suppression of jet yields for a given radius( both small and large) and $p_T$ compared to pp collisions. This has been recently observed in experiments at both Relativistic Heavy Ion Collider (RHIC) \cite{Arsene:2004fa,Back:2004je,Adams:2005dq,Adcox:2004mh} and Large Hadron Collider (LHC) \cite{Aad:2010bu,Aamodt:2010jd,Chatrchyan:2011sx}. The suppression mechanism happens through the mechanism of energy loss when jets travel through the hot medium. There has been tremendous theoretical effort to study the jet energy loss mechanism (see Refs.~\cite{Mehtar-Tani:2013pia,Blaizot:2015lma,Qin:2015srf,Cao:2020wlm} for recent reviews). This means that given a $p_T, R$ bin we are not comparing the evolution of the same hard events in pp versus HIC. One important goal of this paper is to propose observables that are insensitive to jet selection bias allowing us an apples to apples comparison for jet substructure.

The evolution of the jet in the medium usually depends on multiple scales such as the jet energy, the transverse momentum with respect to the jet axis, which will characterize the collinearity of the jet and thermal scales of the QGP. In current heavy ion collision experiments, the temperature achieved lies in the range $150 - 500$ MeV, and may not always be a perturbative scale. Thus, a fully weak coupling calculation may not be valid. A hybrid model has been developed to address this problem \cite{Casalderrey-Solana:2014bpa,Casalderrey-Solana:2015vaa,Hulcher:2017cpt,Casalderrey-Solana:2018wrw,Casalderrey-Solana:2019ubu}, in which the initial jet production and vacuum-like parton shower are calculated perturbatively, while the subsequent jet energy loss in the medium is calculated by mapping the field theory computation in the strong coupling limit to a weak coupling computation in the classical gravity theory \cite{Liu:2006ug,Argyres:2006yz,CasalderreySolana:2007qw,Hatta:2008tx,Chesler:2008uy,DEramo:2010wup}, i.e., by using a modification of the AdS/CFT correspondence \cite{Maldacena:1997re}. 
 
However, since a holographic dual to QCD cannot be rigorously proven, it is unclear whether it can be used for precision jet substructure studies. In this paper, I will instead follow the idea that given a system with multiple hierarchically separated scales, a powerful tool is Effective Field Theory (EFT). The main advantage that an EFT approach offers compared to a purely perturbative Feynman diagram calculation is the notion of factorization, independent of perturbative order. This allows a separation of physics at widely separated scales in terms of manifestly gauge invariant operators which can be independently computed to any perturbative order desired. This approach has been applied with enormous success to  study hadron structure, for e.g. in Deep Inelastic scattering(DIS) where factorization can be used to separate the perturbatively calculable hard function from the non-perturbative Parton Distribution Function(PDF). This establishes an operator definition for a universal hadron structure function which can then be either computed on the lattice or extracted from experiment. 

Such an EFT approach in the context of HIC has been attempted previously in literature using a modification of Soft-Collinear Effective Theory known as SCET$_G$. By making use of the collinear sector of the corresponding  EFT, this formalism has been used to address the question of jet quenching in the medium \cite{Ovanesyan:2011kn,Chien:2015hda,Ovanesyan:2011xy,Chien:2015vja,Kang:2014xsa}. This approach treats the interaction of the jet with the medium in terms of a background gauge field, integrating out the propagating degrees of freedom of the medium.
As a consequence, it is unclear how to establish manifest gauge invariance of operators or prove factorization rigorously in the presence of soft radiation.

I will use an alternative approach applying a new EFT for forward scattering that has been developed recently \cite{Rothstein:2016bsq}. A key difference is that medium degrees of freedom are retained while integrating out the off-shell interaction between the jet and the medium to write manifestly gauge invariant interaction operators. I will show how this can be used to rigorously derive factorization formulas for jet substructure observables and will allow me to provide a operator definition for the QGP medium structure function in analogy with DIS.

The QGP medium exists for a very short time ($\sim 10 fm/c$) so that the the jet spends a limited amount of  time interacting with the medium. At the same time, the medium can be inhomogeneous and evolving with time so that the jet encounters a changing medium as it travels through it. Therefore, we need to keep track of the time evolution of the jet which is a novel feature compared to pp or DIS experiments.
This can most easily be done by using the open quantum systems formalism (for introductory books, see \cite{Breuer:2002pc,OQS}). For jets inside a QGP, if we only focus on jet observables, the jet can be treated as an open quantum system interacting with a QGP bath. The application of the open quantum system formalism in heavy ion collisions has been thriving in the study of color screening and regeneration of quarkonium \cite{Young:2010jq,Borghini:2011ms,Akamatsu:2011se,Akamatsu:2014qsa,Blaizot:2015hya,Katz:2015qja,Kajimoto:2017rel,DeBoni:2017ocl,Blaizot:2017ypk,Blaizot:2018oev,Akamatsu:2018xim,Miura:2019ssi}. The understanding of quarkonium in-medium dynamics has been improved by combining potential nonrelativistic QCD (pNRQCD \cite{Brambilla:1999xf,Brambilla:2004jw,Fleming:2005pd}, an EFT of QCD) and the open quantum system formalism \cite{Brambilla:2016wgg,Brambilla:2017zei,Brambilla:2019tpt,Yao:2018nmy}. For example, a semiclassical Boltzmann transport equation of quarkonium in the medium has been derived, under assumptions that are closely related with a hierarchy of scales \cite{Yao:2018nmy,Yao:2019jir,Yao:2020kqy}.

In this paper, I will combine the tools of EFT with the open quantum system formalism and explore its physical implications for the jet-medium interaction. The long term goal is to develop a theoretically robust formalism for computing jet substructure observables for both light parton and heavy quark jets. For example, the bottom quark jets have been identified as an effective probe of the QGP medium and will be experimentally studied at LHCb, as well as by the sPHENIX collaboration at RHIC. There has been recent work on computing jet substructure observable for heavy quark jets in the context of proton-proton collisions \cite{Lee:2019lge,Makris:2018npl}. The objective would then be to compute the same observables in heavy ion collisions and study modifications caused by the medium.

A first step was taken in \cite{Vaidya:2020cyi} which looked at the transverse momentum spread of a single energetic quark as a function of the time of propagation through the QGP medium. However for a realistic description of the system, we also need to account for the initial hard interaction that creates  the energetic quark which is dressed with radiation from the subsequent parton shower along with any medium interactions. At the same time a realistic final state jet produced in a heavy ion collision will usually have a large number of soft partons that originated from the QGP medium and are not directly associated with the evolution of the energetic jet. Thus any final state measurement imposed on the jet will have to account for these corrections which necessitates keeping track of the degrees of freedom of the QGP medium. A way around this, which has long been used to deal with soft contamination from Multi-Parton Interactions (MPIs) in pp collisions is that of jet grooming (For e.g. Soft-Drop {\cite{Larkoski:2014wba}}), which I ll use in this paper in simplified form.

To implement this, I will borrow the tools developed in the context of pp collision for computing groomed jet substructure observables. This field has progressed rapidly in recent years, both due to advances in explicit calculations, e.g.~\cite{Feige:2012vc}, as well as due to the development of techniques for understanding properties of substructure observables using analytic \cite{Larkoski:2014gra} approaches.  Developments in jet substructure (see \cite{Larkoski:2017jix} for a recent review) have shown that the modified mass drop tagging algorithm (mMDT) or soft-drop grooming procedure robustly removes contamination from both underlying event and non-global color-correlations, see Refs. \cite{Larkoski:2014wba}.

This paper is organized as follows :In Section~\ref{sec:Observable}, I introduce the physical system that we wish to study along with the final state measurements imposed and the relevant physical scales that play an important role in its description. I also describe the relevant momentum modes that are dictated by these scales which will be a guide towards writing down a factorization formula. 
The next Section~\ref{sec:Fact} works out in detail, the factorization formula for the reduced density matrix of the jet within the framework of SCET. Section~\ref{sec:Deco} deals with impact of the lack of coherence between the hard and medium interaction on the factorized density matrix. In Section~\ref{sec:JetMod} I present the one loop results for elastic collisions of the jet with the medium presenting its UV and IR structure. Section~\ref{sec:Master} gives the form of the master evolution equation for multiple jet -medium interactions and solves is analytically. Finally I conclude and discuss future directions in Section~\ref{sec:conclusion}. The details of the loop calculations for the vacuum and medium induced functions are given in Appendix \ref{App:vacuum} and \ref{App:JetM} respectively.


\section{The observable}
\label{sec:Observable}
\begin{figure}
     \centering
     \includegraphics[width=0.6\textwidth]{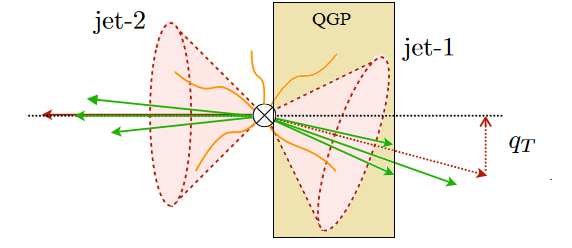}
         \caption{Dijet event in Heavy Ion collision at the periphery of the QGP medium.}
     \label{Obs}
\end{figure}

I want to a consider final state dijet events produced in a heavy ion collision in the background of a QGP medium. The jet are isolated using a suitable jet algorithm such as anti-kT with jet radius $R\sim 1$. We examine the scenario when the hard interaction creating the back to back jets happens at the periphery of the heavy ion collision, so that effectively only one jet passes through the medium while the other evolves purely in vacuum as shown in Fig.\ref{Obs}.
Since we do not want to keep track of the soft partons coming from the QGP, we groom the jets. We put in a simple energy cut-off sufficiently large to remove all partons at energy T and lower. Given a hard scale Q $\sim 2E_J$, where $E_J$ is the energy of the jet and an energy cut-off, $z_{c}E_J$, we work in the hierarchy 
\bea
Q \sim z_{c}Q \gg T 
\eea
where T is the plasma temperature. The measurement we wish to impose is the transverse momentum imbalance between the two jets $q_T \sim T$. 
While this fixes the scaling of all radiation modes that fail grooming, this does not necessarily guarantee collinear scaling inside the jet. To ensure that, we also put in a cumulative jet mass, e, measurement on both jets with $Q\sqrt{e} \sim q_T \sim T$. An identical measurement with $z_{c}<<1$ for $e^+e^- \rightarrow$ dijets was discussed in \cite{Gutierrez-Reyes:2019msa} and we will refer the reader to that paper for a more detailed analysis of this observable.

We wish to write down a factorization theorem within Soft Collinear Effective Theory(SCET) which separates out functions depending on their scaling in momentum space. 
This leads us to the following modes,
\bea 
p_h^{\mu} \sim Q(1, 1, 1), \ \text{Hard function} \nn\\
p_s^{\mu} \sim Q ( \lambda, \lambda, \lambda) , \ \text{Global Soft mode}\nn\\ 
p_n^{\mu} \sim Q(1, \lambda^2, \lambda), \ \ \text{collinear mode}\nn\\
p_{\bar n}^{\mu} \sim Q(\lambda^2, 1, \lambda), \ \ \text{collinear mode}
\eea 
with 
\bea
\lambda = \frac{q_T}{Q} \sim \sqrt{e} \sim \frac{T}{Q}  <<1
\eea
being the expansion parameter of our EFT. 
The medium induces another scale, namely the Debye screening mass $m_D \sim g T$. We will work in a weakly coupled regime so that $m_D$ is a much smaller scale than T. In a completely quantum coherent process such as a pp collision, the imposition of an IRC (Infra-Red collinear) safe observable on the final state guarantees that the final state physics is not sensitive to any IR scale below the one set by the measurement. However, in our case the presence of the medium can induce incoherent scattering  which can potentially make the observable sensitive to $m_D$ as we will see in an explicit calculation.
 
Given the scaling we see that the Global soft mode fails grooming. The emissions that fail grooming lie outside the groomed jet and hence contribute to the the transverse momentum imbalance. The Collinear mode is sensitive to the grooming parameter $z_{c}$ so that it may either pass or fail grooming. When it fails grooming, it will contribute to the transverse momentum imbalance while when it passes, it will contribute to the cumulative jet mass. We assume that the particles that make up the medium also scale uniformly in temperature and have the same scaling as the soft mode. 

From the scaling of the modes, we see that the constraints from the observable imposed imply that there is no radiation that is sensitive to the edge of the jet and hence to the jet radius. The soft modes all fail grooming and lie outside the groomed jet while the collinear radiation is confined to a narrow core near the jet axis. Hence, there are no sources of energy/ transverse momentum leakage near the edge of the jet. There will be jet substructure modification $within$ the jet without changing its radius and $pT$ so that it now becomes possible to compare the jet substructure modification for the same hard events in pp and HIC even while working in the same $p_T$, R bin.


\section{Factorization for Reduced density matrix evolution}
\label{sec:Fact}
I would like to derive a factorization formula for this observable showing a clear separation of scales in terms of gauge invariant operators.  I will treat the jet as an open quantum system interacting with the QGP bath and follow the time evolution of the reduced density matrix of the jet. Along the way, I will work out the factorization for this density matrix within the framework of SCET.

For ease of analysis, we consider that the hard interaction that creates the jet is an $e^+e^-$ collision. While this is not a real scenario, it is an ideal playground to work out the EFT framework which mainly deals with the final state physics. The EFT structure can then be easily carried over to the realistic case of nuclear/hadronic collisions, which we leave for future analysis as part of a detailed phenomenological application of the formalism developed in this paper. While a sketch of the factorization for the same observable with a different hierarchy of scales was provided in \cite{Gutierrez-Reyes:2019msa}, we revisit the factorization in the context of a density matrix evolution, which uses time ordered perturbation theory. \\
The hard interaction can be encoded using an effective current operator 
\bea
\mathcal{O}_{H} =  C(Q)L^{\mu} J^{SCET}_{\mu} 
\eea
$C(Q)$ is the Wilson co-efficient for this contact operator that depends only on the hard scale Q. This will lead to a hard function H(Q) at the amplitude squared level and its form is discussed in Appendix \ref{App:vacuum}. 
$L^{\mu}$ is the initial state current, which in this case is just the lepton current, while $J^{SCET}_{\mu} $ would be the final state SCET current, which is just the gauge invariant quark current. 
\bea
L^{\mu} = \bar l \gamma^{\mu} l, \ \ \  J^{SCET}_{\mu}= \bar \chi_n \gamma^{\mu} \chi_{\bar n} \nn
\eea
where $ n =(1,0,0,1)$, and $\bar n = (1,0,0,-1)$ are light-like vectors pointing in the direction of the initial back to back $q \bar q$ pair .
 The initial state density matrix would then be 
\bea
\rho(0) = |e^+ e^-\rangle \langle e^+e^-| \otimes \rho_B
\eea 
 Since the partons in the medium have the same scaling as the soft mode, we will henceforth suppress explicitly writing out the factor  $\rho_B$ till it becomes relevant for the Soft function analysis. We have started with the assumption that the initial state participating in the hard interaction is disentangled from the state of the background medium.
 We can follow the evolution of this density matrix which will evolve with the effective Hamiltonian 
\bea
H=H^{\text{IR}}+ \mathcal{O}_{H}
\eea

Since I am interested in a dijet event, I  will consider only a single insertion of the hard operator which, at tree level will create back to back  $q \bar q$ pair that subsequently evolve into dijets. The IR  Hamiltonian consists of the Hamiltonians that describe the IR modes and interactions between them including the Glauber Hamiltonian that mediates forward interaction between the jet and the medium.  As motivated in \cite{Vaidya:2020cyi}, the dominant interaction of the jet with the medium in mediated by the t channel exchange of the Glauber gluon. The detailed form of $H^{\text{IR}}$ will be discussed later in this section. 
The time evolved density matrix is given as 
\bea
\rho(t)&=& e^{-iHt}\rho(0)e^{iHt} = e^{-iH^{\text{IR}}t}\Big[e^{iH^{\text{IR}}t} e^{-iHt}\Big]\rho(0) \Big[e^{iHt}e^{-iH^{\text{IR}}t}\Big] e^{iH^{\text{IR}}t} \nn\\
&=& e^{-iH^{\text{IR}}t}U(t,0)\rho(0)U^{\dagger}(t,0) e^{iH^{\text{IR}}t}
\eea
Our evolution operator $U(t,0)$ now obeys the equation 
\bea
\partial_t U(t,0)= -i[\mathcal{O}_{H,I}(t),U(t,0)],  \ \ \  \text{with} \ \ \  \mathcal{O}_{H,I}(t)= e^{iH^{\text{IR}}t}\mathcal{O}_{H}e^{-iH^{\text{IR}}t}
\eea
which has the solution 
\bea
U(t,0) = \mathcal{T} \Big\{ e^{-i\int_0^t dt' \mathcal{O}_{H,I}(t')} \Big\} 
\eea
which is the evolution operator written as a time ordered exponent of the dressed hard operator. 
Our solution for the density matrix now becomes 
\bea
 \rho(t)=  e^{-iH^{\text{IR}}t}\mathcal{T} \Big\{ e^{-i\int_0^t dt' \mathcal{O}_{H,I}(t')} \Big\} \rho(0) \bar{\mathcal{T}} \Big\{ e^{-i\int_0^t dt' \mathcal{O}_{H,I}(t')} \Big\} e^{iH^{\text{IR}}t}
\eea
We are interested in creating dijets, it is sufficient to consider a single insertion of the Hard operator on each side of the cut. 
\bea
 \rho(t)= e^{-iH^{\text{IR}}t}\rho(0)e^{iH^{\text{IR}}t}+ \int_0^t dt_1\int_0^t dt_2e^{-iH^{\text{IR}}t}\mathcal{O}_{H,I}(t_1)\rho(0) \mathcal{O}^{\dagger}_{H,I}(t_2)e^{iH^{\text{IR}}t}
\eea
When we impose the measurement for the dijet event with the required properties, only the second terms will survive, hence hereafter we can simply follow the evolution for this piece.

We define 
\bea
\sigma(t) &=&   \int_0^t dt_1\int_0^t dt_2e^{-iH^{\text{IR}}t}\mathcal{O}_{H,I}(t_1)\rho(0) \mathcal{O}^{\dagger}_{H,I}(t_2)e^{iH^{\text{IR}}t}\nn\\
&=& |C(Q)|^2I^{\mu\nu}\int d^3x_1\int_0^t dt_1\int d^3x_2\int_0^t dt_2 e^{-i(x_1-x_2)\cdot(p_e+p_{\bar e})}\nn\\
&&e^{-iH^{\text{IR}}t}J^{\mu}_{SCET}(x_1)|0\rangle \langle 0| J^{\nu}_{SCET}(x_2)e^{iH^{\text{IR}}t}
\eea
where $p_e$, $p_{\bar e}$ are the momenta of the initial state electron positron. In the c.o.m. frame $p_e+p_{\bar e} =(Q, 0, 0, 0)$.
$I^{\mu \nu}$ is the Lepton tensor. The IR Hamiltonian is written as a sum over the Hamiltonians of all the SCET sectors( i.e., the n collinear, $\bar n$ collinear,  Soft) along with the Glauber Hamiltonian, which in our case introduces a interaction between the Soft and  n collinear sectors. 
\bea 
H^{\text{IR}} = H^{SCET}+H^G =  H_n+H_{\bar n}+H_S + H^G_{ns}
\eea 
This form of the QCD Hamiltonian is correct to leading power in our expansion parameter $\lambda$. The Glauber Hamiltonian is expressed in terms of effective gauge invariant operators for quark-quark ($qq$), quark-gluon ($qg$ or $gq$) and gluon-gluon ($gg$) interactions which have been worked out in the Feynman gauge in Ref.~\cite{Rothstein:2016bsq}
\bea
\label{EFTOp}
H^G &=& \sum_{ij} C_{ij}\mathcal{O}_{ns}^{ij} \nn\\
\mathcal{O}_{ns}^{qq}&=&\mathcal{O}_n^{qB}\frac{1}{\mathcal{P}_{\perp}^2}\mathcal{O}_s^{q_nB} , \ \ \ \mathcal{O}_{ns}^{qg}=\mathcal{O}_n^{qB}\frac{1}{\mathcal{P}_{\perp}^2}\mathcal{O}_s^{g_nB} , \nn\\
\mathcal{O}_{ns}^{gq}&=&\mathcal{O}_n^{gB}\frac{1}{\mathcal{P}_{\perp}^2}\mathcal{O}_s^{q_nB} , \  \ \  \mathcal{O}_{ns}^{gg}=\mathcal{O}_n^{gB}\frac{1}{\mathcal{P}_{\perp}^2}\mathcal{O}_s^{g_nB}
\eea
where $B$ is the color index and the subscripts $n$ and $s$ denote the collinear and soft operators. The Glauber gluon propagator appears as the derivative in $\perp$ direction. We will assume for the remainder of the paper that the jet which traverses the medium points along the n direction. $C_{ij}$ are the Wilson co-efficients for these contact operators and all begin at $O(\alpha_s)$.

The SCET current is given as 
\bea
J^{\mu}_{SCET}= \Big[S_n^{\dagger}S_{\bar n}\Big]\bar \chi_n W_n^{\dagger} \gamma^{\mu} W_{\bar n}\chi_{\bar n}
\label{JSCET}
\eea

where $S_i, W_j$ are soft and collinear Wilson lines, defined as 
\bea
S_n^{(r)}(x) = \text{P} \exp \Bigg[ig \int_{-\infty}^0 ds n \cdot A_s^B(x+sn)T^B_{(r)}\Bigg]\nn\\
 W_n(x)=\Bigg[\sum_{\text{perms}} exp\left(-\frac{g}{n\cdot\mathcal{P}}\bar n\cdot A_n(x)\right)\Bigg]
\eea
We see that the SCET current is already factorized in terms of Soft and Collinear sectors which are mainifestly gauge invariant. At the same time they are decoupled from each other in $H_{SCET}$. However, the  Glauber Hamiltonian prevents us from factorizing the full density matrix since it couples the collinear n and the Soft sectors. We therefore need to expand in powers of the Glauber Hamiltonian and establish factorization at each order in the Glauber expansion. We will show that after imposing final state measurements, it is possible to resum  the series in the Glauber Hamiltonian, atleast in the Markovian approximation via a Lindblad type equation.

To proceed, we rearrange the the result above so as to be able to do a systematic expansion in the Glauber Hamiltonian. 
For convenience lets define 
\bea
\int d \tilde x =  \int d^3x_1\int_0^t dt_1\int d^3x_2\int_0^t dt_2e^{-i(x_1-x_2)\cdot(p_e+p_{\bar e})}
\eea
\bea
\sigma(t) &=& |C(Q)|^2 I^{\mu\nu}\int d \tilde x e^{-iH^{\text{IR}}t}J^{\mu}_{SCET}(x_1)|0\rangle \langle 0| J^{\nu}_{SCET}(x_2)e^{iH^{\text{IR}}t}\nn\\ 
&=& |C(Q)|^2 I^{\mu\nu}\int d \tilde x e^{-iH^{SCET}t}\Big\{e^{iH^{SCET}t}e^{-iH^{\text{IR}}(t-t_1)}e^{-iH^{SCET}t_1}\Big\} \nn\\
&&\Big\{e^{iH^{SCET}t_1}J^{\mu}_{SC}(\vec{x}_1,0)e^{-iH^{SCET}t_1}\Big\}\Big\{e^{iH^{SCET}t_1}e^{-iH^{\text{IR}}t_1}\Big\}|0\rangle \langle 0| \Big\{e^{iH^{\text{IR}}t_2}e^{-iH^{SCET}t_2}\Big\}\nn\\
&&\Big\{e^{iH^{SCET}t_2}J^{\nu}_{SCET}(\vec{x}_2,0)e^{-iH^{SCET}t_2}\Big\}\Big\{e^{iH^{SCET}t_2}e^{iH^{\text{IR}}(t-t_2)}e^{-iH^{SCET}t}\Big\}e^{iH^{SCET}t}\nn
\eea
Using the same process as for the hard operator, we can rearrange the expression as time ordered dressed operators 
\bea
 \sigma(t)&=&  |C(Q)|^2 I^{\mu\nu}\int d \tilde x e^{-iH^{SCET}t}\mathcal{T} \Big\{ e^{-i\int_0^t dt' H_{G,I_{SC}}(t')} J^{\mu}_{SCET,I_{SC}}(x_1)\Big\} |0\rangle \nn\\
&& \langle 0| \bar{\mathcal{T}} \Big\{ e^{-i\int_0^t dt' H_{G,I_{SC}}(t')}J^{\nu}_{SCET,I_{SC}}(x_2) \Big\} e^{iH^{SCET}t}
\label{GExp}
\eea
where 
\bea
O_{I_{SC}}(t) =  e^{iH^{SCET}t}Oe^{-iH^{SCET}t} 
\eea
so that all operators are now dressed with the SCET Hamiltonian. We are now set up to do an expansion in the Glauber Hamiltonian. Ultimately, for this paper, we want to compute the trace over the reduced density matrix with an appropriate measurement 
\bea
\label{Sigm}
\Sigma(t) \equiv \text{Tr}[\sigma(t) \mathcal{M}]\big|_{t\rightarrow \infty} 
\eea
where we first completely trace over the Soft and collinear degrees of freedom with the measurement $\mathcal{M}$ which includes the jet algorithm to isolate a final state large radius groomed dijet configuration with the required $q_T$ imbalance and jet mass. 
We then expand this out in powers of $H_G$
\bea
\label{LEQ}
 \Sigma(t) = \Sigma^{(0)}(t) + \Sigma^{(1)}(t)+ \Sigma^{(2)}(t)+ O(H_G^3)+...
\eea
In the next section we will sketch the proof for factorization of the reduced density matrix upto quadratic order in the $H^G$ expansion. We will subsequently use this to derive a Lindblad equation to resum all the higher order terms in $H^G$ in the Markovian approximation.

\subsection{Leading order in Glauber:Vacuum evolution}

We start with the leading order term from Eq.\ref{GExp} with no Glauber insertions and so should simply give us a result proportional to the vacuum-background cross section. Since we are doing this in the context of time ordered perturbation theory we will outline the proof for factorization of this piece here.
\bea
\sigma^{(0)}(t) &=&  |C(Q)|^2I^{\mu\nu}\int d \tilde x e^{-iH_{SCET}(t-t_1)} J^{\mu}_{SCET}(\vec{x}_1)e^{-iH^{SCET}t_1} |0\rangle \nn\\
&&\langle 0|e^{iH^{SCET}t_2} J^{\nu}_{SCET}(\vec{x}_2)e^{iH^{SCET}(t-t_2)}\nn
\eea

We can now write the result in the interaction picture separating out the free theory Hamiltonian $H_0$ from the interactions $H_{int}$,
\bea
H^{SCET} = H_0+ H_{int} 
\eea
then performing the same series of steps as before, we can write 
\bea
\sigma^{(0)}(t) &=&|C(Q)|^2I^{\mu\nu}\int d \tilde x e^{-iH_{0}t}\mathcal{T} \Big\{ e^{-i\int_0^t dt' H_{int,I}(t')} J^{\mu}_{SCET,I}(x_1)\Big\} |0\rangle \nn\\
&&\langle 0| \bar{\mathcal{T}} \Big\{ e^{-i\int_0^t dt' H_{int,I}(t')}J^{\nu}_{SCET,I}(x_2) \Big\} e^{iH_{0}t}\nn
\eea
where 
\bea
O_{I}(t) =  e^{iH_{0}t}Oe^{-iH_{0}t} 
\eea
To proceed further, we put in our measurement on the dijets and take a trace over final states and take the limit $t \rightarrow \infty$.
\bea
&&\langle X| \sigma(t \rightarrow \infty)\mathcal{M}|X\rangle \equiv \Sigma^{(0)}= |C(Q)|^2I^{\mu\nu}\int d \tilde x  \nn\\
&&\langle X|\mathcal{T} \Big\{ e^{-i\int_0^{\infty} dt' H_{int,I}(t')} J^{\mu}_{SCET,I}(x_1)\Big\} |0\rangle \langle 0| \bar{\mathcal{T}} \Big\{ e^{-i\int_0^{\infty} dt' H_{int,I}(t')}J^{\nu}_{SCET,I}(x_2) \Big\}\mathcal{M}|X\rangle  \nn
\eea
where we have 
\bea
\mathcal{M} = \delta^{2}\left( \vec{q}_T- \vec{p}^{\perp}_{n,\in \text{gr}}-\vec{p}^{\perp}_{\bar n, \in \text{gr}}\right)\Theta\left(e_n- e_{n,\in \text{gr}}\right) \Theta \left(e_{\bar n}-e_{\bar n,\in \text{gr}}\right) \nn
\eea
where $\vec{q}_T$ is the transverse momentum imbalance between the two groomed jets, while $e_n$, $e_{\bar n}$ measures the jet mass for the two groomed jets. The jet mass is defined as 
\bea
e_{n,\in \text{gr}} = \frac{4}{Q^2}\left(\sum_{i\in n,\text{gr}}p_i \right)^2 
\eea
Notice that none of the modes are sensitive to the jet radius R. This follows from the fact that the collinear radiation is collimated along the jet axis far away from the edge of the jet while the soft radiation can populate the full phase space without constraint. 
Our Hamiltonian is
\bea
H_{int} =  H_S+ H_n +H_{\bar n} 
\eea
where the three sectors are decoupled from each other while the interactions between the various sectors now appear in the form of  Wilson lines in the SCET current (Eq. \ref{JSCET}). Without the presence of factorization violating Glauber interactions, using standard techniques outlined in \cite{Gutierrez-Reyes:2019msa}, the momentum sectors can now be written as separate matrix elements. Hence the Hilbert space also factorizes into various momentum mode states
\bea
 |X\rangle = |X_n\rangle |X_{\bar n}\rangle |X_s \rangle 
\eea 
Since the Hamiltonian is already factorized, we have, in principle, a factorization of all the modes at this stage. 
However, we still have to implement  our power counting on the measurement functions which will ensure that only leading power corrections in our expansion parameter $\lambda$ are retained. \\
Acting on the final state Hilbert space, we can then pull out the co-ordinate dependence of the SCET current and perform all co-ordinate integrals 
\bea
\Sigma^{(0)}&=& |C(Q)|^2I^{\mu \nu} \int d^4x_1 \int d^4x_2 e^{-i(x_1-x_2)\cdot(p_e+p_{\bar e}-p_{X_n}-p_{X_{\bar n}}-p_{X_s})}\nn\\
&\times& \langle X|\mathcal{T} \Big\{ e^{-i\int_0^{\infty} dt' H_{int,I}(t')} J^{\mu}_{SCET,I}(0)\Big\} |0\rangle \langle 0| \bar{\mathcal{T}} \Big\{ e^{-i\int_0^{\infty} dt' H_{int,I}(t')}J^{\nu}_{SCET,I}(0) \Big\}\mathcal{M}|X\rangle \nn
\eea
Performing the integrals over $x_1$ and $x_2$ now gives momentum conserving $\delta $ function along with a 4d volume factor V. We can then decompose the 4 momentum delta function in light-cone co-ordinates and apply power counting 
\bea
&&\delta^4(p_e+p_{\bar e}-p_{X_n}-p_{X_{\bar n}}-p_{X_s}) \nn\\
&\rightarrow& \delta(Q- p_{X_n}^-)\delta(Q-p^+_{X_{\bar n}})\delta^2(p^{\perp}_{X_{n, \in \text{gr}}}+p^{\perp}_{X_{\bar n,\in \text{gr}}}+p^{\perp}_{X_s}+p^{\perp}_{X_{n,\not\in \text{gr}}}+p^{\perp}_{X_{\bar n,\not\in \text{gr}}})
\eea
where the subscript $\not\in \text{gr}$ indicates collinear radiation that fails the grooming condition. Our factorization now becomes 
\bea
 \Sigma^{(0)}&=&V\times |C(Q)|^2I^{\mu \nu} \langle X|\mathcal{T} \Big\{ e^{-i\int_0^{\infty} dt' H_{int,I}(t')} J^{\mu}_{SCET,I}(0)\Big\} |0\rangle \nn\\
&&\langle 0| \bar{\mathcal{T}} \Big\{ e^{-i\int_0^{\infty} dt' H_{int,I}(t')}J^{\nu}_{SCET,I}(0) \Big\}|X\rangle \nn\\
&\times& \delta(Q- p_{X_n}^-)\delta(Q-p^+_{X_{\bar n}})\delta^2(p^{\perp}_{X_{n,\in \text{gr}}}+p^{\perp}_{X_{\bar n,\in \text{gr}}}+p^{\perp}_{X_s}+p^{\perp}_{X_{n,\not\in \text{gr}}}+p^{\perp}_{X_{\bar n,\not\in \text{gr}}}) \nn\\
&\times&\delta^{2}\left( \vec{q}_T- \vec{p}^{\perp}_{n,\in \text{gr}}-\vec{p}^{\perp}_{\bar n, \in \text{gr}}\right)\Theta \left(e_n- e_{n,\in \text{gr}}\right) \Theta \left(e_{\bar n}-e_{\bar n,\in \text{gr}}\right) \nn
\eea
Without loss of generality, we can assume that the axis of the $\bar n$ groomed jet is exactly aligned with the $ \bar n$ direction, in which case its transverse momentum becomes zero.
\bea
\Sigma^{(0)}&=&V\times |C(Q)|^2I^{\mu \nu} \langle X|\mathcal{T} \Big\{ e^{-i\int_0^{\infty} dt' H_{int,I}(t')} J^{\mu}_{SCET,I}(0)\Big\} |0\rangle \nn\\
&&\langle 0| \bar{\mathcal{T}} \Big\{ e^{-i\int_0^{\infty} dt' H_{int,I}(t')}J^{\nu}_{SCET,I}(0) \Big\}\mathcal{M}|X\rangle \nn\\
&\times& \delta(Q- p_{X_n}^-)\delta(Q-p^+_{X_{\bar n}})\delta^2(\vec{q}_T+p^{\perp}_{X_s}+p^{\perp}_{X_{n,\not\in \text{gr}}}+p^{\perp}_{X_{\bar n,\not\in \text{gr}}}) \delta^2_{\mathcal{P}^{\perp}_{\bar n}}\nn\\
&\times&\delta^{2}\left(\vec{q}_T- \vec{p}^{\perp}_{n,\in \text{gr}}\right)\Theta\left(e_n- e_{n,\in \text{gr}}\right) \Theta\left(e_{\bar n}-e_{\bar n,\in \text{gr}}\right) \nn
\eea
$\delta_{\mathcal{P}^{\perp}_{\bar n}}$ is a Kronecker delta setting the transverse momentum of the $\bar n$ groomed jet to 0. 
We now have a transverse momentum condition $\delta^{2}\left( \vec{q}_T- \vec{p}^{\perp}_{n, \in \text{gr}}\right)$, which tells us that the n groomed  jet is not exactly aligned with the n axis. However, we can use RPI I invariance of from SCET, to adjust the axis of the groomed jet without changing any physics, so that this condition can simply to written as $\delta^2( \vec{p}^{\perp}_{n,\in \text{gr}})$
which now gets us back to the standard definition of the jet function. We also see that the transverse momentum imbalance receives contributions from all the modes that fail grooming. \\
We can convert the Kronecker delta to a Dirac delta following literature \cite{Chiu:2012ir}
\bea
\delta^2_{\mathcal{P}^{\perp}_{\bar n}}= \pi Q^2\delta^2\left(p^{\perp}_{X \bar n,\in \text{gr}}\right)
\eea

The jet mass measurement receives contributions from the collinear modes that pass grooming.
We can now write the final form of our factorized density matrix element
\bea
\label{SigmZ}
\Sigma^{(0)}(q_T,e_n,e_{\bar n}) =V \times H(Q,\mu)S(\vec{q}_T;\mu,\nu)\otimes_{q_T}\mathcal{J}^{\perp}_n(e_n,Q,z_{c},\vec{q}_T;\mu,\nu)\otimes_{q_T} \mathcal{J}^{\perp}_{\bar n}(e_{\bar n},Q,z_{c},\vec{q}_T;\mu,\nu)\nn\\
\eea
where $\otimes_{q_T}$ indicates a convolution in $\vec{q}_T$.
H(Q) is the hard function which also includes the born level term. The factorized functions are defined as follows 
\bea
\label{eq:soft}
 S(\vec{q}_T) = \frac{1}{N_R}\text{tr}  \langle X_S| \mathcal{T}\Big\{e^{-i\int_0^{\infty}dt' H_S(t')}S_{\bar n}^{\dagger}S_n(0)\Big\}|0\rangle \langle 0| \mathcal{\bar T}\Big\{e^{-i\int_0^{\infty}dt' H_S(t')}S_{n}^{\dagger}S_{\bar n}(0)\Big\}\delta^2(\vec{q}_T-\mathcal{P}_{\perp})|X_S\rangle \nn\\
\eea
The trace here is a trace over color and its understood that $|X_S\rangle \langle X_S|$ includes a sum over soft states with their phase space integrated over. This computes the Soft function in a vacuum background, but as we know we actually have a background of the medium particles which also scales as the soft mode. So, in principle, we have
\bea
  S(\vec{q}_T) =  \frac{1}{N_R}\text{tr}  \langle X_S| \mathcal{T}\Big\{e^{-i\int_0^{\infty}dt' H_S(t')}S_{\bar n}^{\dagger}S_n(0)\Big\}\rho_{B} \mathcal{\bar T}\Big\{e^{-i\int_0^{\infty}dt' H_S(t')}S_{n}^{\dagger}S_{\bar n}(0)\Big\}\delta^2(\vec{q}_T-\mathcal{P}_{\perp})|X_S\rangle \nn
\eea
where we have assumed a time independent QGP background. Of course, we can take into account the fact that the time scales for the soft emission( which puts the collinear mode off-shell) is much shorter than the formation time for the QGP, in which case we would be justified to compute the soft function in a vacuum background, which is what we will assume for the rest of this paper. The quark jet function is defined as 
\bea
\mathcal{J}_n^{\perp}(e,Q) &=& \frac{(2\pi)^3}{N_c} \text{tr}\langle X_n| \mathcal{T}\Big\{e^{-i\int_0^{\infty}dt' H_n(t')}\bar \chi_n(0)\Big\}|0\rangle\nn\\
&& \langle 0|\mathcal{\bar T}\Big\{e^{-i\int_0^{\infty}dt' H_n(t')}\frac{\slashed{\bar n}}{2}\chi_n\Big\}\delta(Q-\mathcal{P}^-)\delta^2(\mathcal{P}^{\perp})\Theta(e_n- \mathcal{E}_{\in n,\text{gr}})\delta^2(\vec{q}_T-\mathcal{P}^{\perp}_{\not\in n,\text{gr}})|X_n\rangle  \nn
\eea
  The Renormalization group equation for each function can be solved in impact parameter space where the convolution in $\vec{q}_T$ turns into a product 
\bea
 \Sigma^{(0)} =V \times H(Q,\mu)\times \int d^2b e^{i \vec{q}_{T}\cdot \vec{b}}S(\vec{b};\mu,\nu)\mathcal{J}^{\perp}_n(e_n,Q,z_{cut},\vec{b} ;\mu, \nu) \mathcal{J}^{\perp}_{\bar n}(e_{\bar n},Q,z_{cut},\vec{b};\mu,\nu)\nn\\
\eea

where for a function $F(\vec{q}_T)$,
\bea
F(\vec{b}) = \int \frac{d^2\vec{q}_T}{(2\pi)^2}e^{-i \vec{q}_T \cdot \vec{b}}F(\vec{q}_T)
\eea
The one loop results for all the functions in $\vec{b}$ space are presented in Appendix \ref{App:vacuum} along with resummation.

\subsection{Next-to-Leading order in Glauber}

We now consider the next to leading order term in the expansion of the Glauber Hamiltonian starting from Eq.\ref{Sigm}. In principle, we should start off with a single insertion of $H_G$ on either side of the cut. Since we have a non-trivial soft function consisting  of Soft Wilson lines (Eq. \ref{eq:soft}) at leading order which scales the same way as the soft partons of the medium, the single $H_G$ insertion can lead to interfering diagrams between the soft Wilson lines and the medium. However, as stated in the previous section, if the time scale for Soft emissions is shorter than the QGP formation time, then we can factorize the QGP interactions of the jet from the explicit soft radiation off the quark created in the hard interaction in which case we need to do atleast a quadratic Glauber insertion. This also follows another approximation we will make in Section \ref{sec:Deco}, where we retain only the diagrams where the partons created in the hard interaction go on-shell before interacting with the medium. 
With a quadratic Glauber insertion, we can have two contributions depending on whether the two Glauber insertions are on the same or opposite sides of the cut. This respectively corresponds to a single virtual and real interaction of the jet with the medium.
\bea
\Sigma^{(2)}(t) =  \Sigma_R^{(2)}(t)+\Big\{\Sigma_V^{(2)}(t)+c.c. \Big\}
\eea

\begin{itemize}
\item{ {\bf Glauber insertion on both sides of the cut}}

\bea
\sigma_{R}^{(2)}(t) &=& |C(Q)|^2I^{\mu \nu}\int d\tilde x e^{-iH^{SCET}t}\mathcal{T}\Big\{\int_0^t dt'H_{G,I_{SC}}(t')J^{\mu}_{SCET,I_{SC}}(x_1)\Big\}|0\rangle \nn\\
&&\langle 0|\mathcal{\bar T}\Big\{\int_0^t d\hat t H_{G,I_{SC}}(\hat t)J^{\nu}_{SCET,I_{SC}}(x_2)\Big\}e^{iH^{SCET}t}
\eea

By following the same series of steps as for the leading order term, we can write the result in terms of the free theory interaction picture.
\bea
\sigma_{R}^{(2)}(t) &=&|C(Q)|^2I^{\mu \nu}\int d\tilde x e^{-iH_{0}t}\mathcal{T}\Big\{e^{-i\int_0^{t}dt' H_{int,I}(t')}\int_0^t dt_a H_{G,I}(t_a)J^{\mu}_{SCET,I}(x_1)\Big\}|0\rangle \nn\\
&\times& \langle 0|\mathcal{\bar T}\Big\{e^{-i\int_0^{t}dt' H_{int,I}(t')}\int_0^t dt_b H_{G,I}(t_b)J^{\nu}_{SCET,I}(x_2)\Big\}e^{iH_0t}\nn
\eea

The next step is to obtain a factorized formula in terms of our EFT modes. To do this, we explicitly put in the form of our Glauber operator,  considering the case of collinear partons interacting with the soft partons in the medium. These operators were defined in Eq. \ref{EFTOp}
\bea
H_{G}(t) =\sum_{i,j \in {q,g}} C_{ij} \int d^3 \vec{x} O_{n,i}^A(\vec{x},t) \frac{1}{\mathcal{P}_{\perp}^2}O_{S,j}^A(\vec{x},t) 
\eea

We can now take the trace over the density matrix inserting our measurement as before. 
\bea
&&\langle X |\mathcal{M} \sigma_R^{(2)}(t\rightarrow \infty)|X \rangle \equiv \Sigma^{(2)}_R \nn\\
&=& |C(Q)|^2I^{\mu \nu}\int d\tilde x
\langle X|\mathcal{T}\Big\{e^{-i\int_0^{t}dt' H_{int,I}(t')}\int_0^t dt_a H_{G,I}(t_a)J^{\mu}_{SCET,I}(x_1)\Big\}|0\rangle \nn\\
&\times& \langle 0|\mathcal{\bar T}\Big\{e^{-i\int_0^{t}dt' H_{int,I}(t')}\int_0^t dt_b H_{G,I}(t_b)J^{\nu}_{SCET,I}(x_2)\Big\}\mathcal{M}|X\rangle \nn
\eea
We can now follow the same series of steps as for the leading order term, and apply power counting to measurement functions as well as the momentum conserving $\delta$ functions based on the momentum scaling of each mode. Accordingly, we factorize the Hilbert space of the final states in terms of the momentum scaling of the modes and pull out the co-ordinate dependence of each mode by acting with the operators on the final state. This yields the following co-ordinate integrals
\bea
I&=&\int d^4x_1 e^{-ix_1\cdot(p_e+p_{\bar e}-p_{Jn,1}-p_{JS,1}-p_{J\bar n})}\int d^4x_2 e^{-ix_2\cdot(p_e+p_{\bar e}-p_{Jn,2}-p_{JS,2}-p_{J\bar n})} \nn\\
&&\int d^4x_a e^{-ix_a \cdot (p_{Gn,1}+p_{Gs,1})}\times\int d^4x_b e^{-ix_b \cdot (p_{Gn,2}+p_{Gs,2})}
\eea
where the subscripts G, J tell us whether the momentum is coming from the action of the Glauber Hamiltonian fields or the SCET current respectively. 
Now performing the integrals and applying power counting, we have 
\bea
I&=&\delta(Q-p^-_{Jn1})\delta(Q-p_{\bar n}^+)\delta^2(p^{\perp}_{Jn,1}+p^{\perp}_{Js,1}+p^{\perp}_{\bar n})\nn\\
&& \delta(Q-p^-_{Jn2})\delta(Q-p_{\bar n}^+)\delta^2(p^{\perp}_{Jn,2}+p^{\perp}_{Js,2}+p^{\perp}_{\bar n})\nn\\
&&\delta(p^-_{Gn,1})\delta(p^+_{GS,1}+p^+_{Gn,1})\delta^2(p^{\perp}_{Gn,1}+p^{\perp}_{GS,1})\nn\\
&&\delta(p^-_{Gn,2})\delta(p^+_{GS,2}+p^+_{Gn,2})\delta^2(p^{\perp}_{Gn,2}+p^{\perp}_{GS,2})
\eea
This simplification follows from the idea that $p_{GS}$ scales as the Glauber momentum. We have ignored any factors of $2\pi$ which will be absorbed in the overall co-efficient for $\Sigma_R^{(2)}$. 
We also have additional constraints since the total momentum for a particular mode must match on both sides of the cut 
\bea
p_{Jn,1}+p_{Gn,1}= p_{Jn,2}+p_{Gn,2}\nn\\
p_{JS,1}+p_{GS,1}= p_{JS,2}+p_{GS,2}
\eea
We can simplify our measurement $\delta$ functions using these set of constraints 
\bea
&&\delta^2(\vec{q}_T- p^{\perp}_{n,\in \text{gr}}-p^{\perp}_{\bar n, \in \text{gr}})\equiv \delta^2(\vec{q}_T- p^{\perp}_{n,\in \text{gr}}-p^{\perp}_{\bar n, \in \text{gr}})\nn\\
&=&  \delta^2(\vec{q}_T- p^{\perp}_{Gn,1 \in \text{gr}}- p^{\perp}_{Jn,1 \in \text{gr}}-p^{\perp}_{\bar n,\in \text{gr}})\nn\\
&=& \delta^2(\vec{q}_T +[p^{\perp}_{GS,1}+p^{\perp}_{JS,1}]+p^{\perp}_{Jn,1\not\in \text{gr}}+p^{\perp}_{\bar n,\not\in \text{gr}})
\eea
where the term inside the square brackets is the total contribution from the Soft sector, which includes both the vacuum as well as medium effects.
As for the leading order term, we can set the axis of the $\bar n$ jet  to be exactly aligned with the $\bar n$ axis  and then using RPI I, do the same for the n jet before it interacts with the medium. Using the rest of the constraints then, once again we have an overall factor of $V$. Since we are ignoring interference between the Soft operators of the SCET and the Glauber insertion, we can set 
\bea
p_{GS,1}= p_{GS,2} \equiv p_{GS}, \ \ \text{so that} \ \ p_{JS,1}= p_{JS,2}.\nn
\eea

We can now write down our factorization formula for $\Sigma_R^{(2)}$, explicitly pulling out the vacuum Soft function, 
\bea
\Sigma^{(2)}_R &=& V\times |C_{G}|^2 H(Q,\mu) \Big\{\int d^2\vec{q}_{JS} S(\vec{q}_{JS})\Big\} \Big\{ \int d^2\vec{q}_{GS} dp_{GS}^+ S_{G}^{AB}( q^{\perp}_{GS},p_{GS}^+)\Big\}\nn\\
&&\Big\{\int d^2\vec{q}_n\int d^2\vec{p}^{\perp}_{Gn,1}d^2\vec{p}^{\perp}_{Gn,2} dp_{Gn,1}^+ dp_{Gn,2}^+ J_{n}^{AB}(e_n,\vec{p}^{\perp}_{Gn,1},\vec{p}^{\perp}_{Gn,2}, p_{Gn,1}^+, p_{Gn,2}^+,\vec{q}_n)\Big\}\nn\\
&\times& \int d^2\vec{q}_{\bar n}J_{\bar n}(e_{\bar n},\vec{q}_{\bar n})\delta^2(\vec{q}_T+ \vec{q}_{JS}+\vec{q}_{GS}+\vec{q}_{n}+\vec{q}_{\bar n})\nn\\
&\times& \delta^2(\vec{q}_{GS}+\vec{p}^{\perp}_{Gn,1})\delta^2(\vec{q}_{GS}+\vec{p}^{\perp}_{Gn,2})\delta(p_{GS}^++ p_{Gn,1}^+)\delta( p_{GS}^++p_{Gn,2}^+)
\eea

In order to simplify notation, its easiest to express this result by rewriting some of the momentum conserving $\delta$ functions in co-ordinate space. This gives us 
\bea
\Sigma^{(2)}_R &=& V\times|C_G|^2H(Q,\mu) \Big\{\int d^2\vec{q}_{JS} S(\vec{q}_{JS})\Big\}\nn\\
&\times&\int d^4x \int d^4 y \Big\{ \int d^2\vec{q}_{GS}\bar{S}_G^{AB}( q^{\perp}_{GS},\{x_{\perp},x^-\},\{y_{\perp},y^-\})\Big\}\Big\{\int d^2\vec{q}_n J_n^{AB}(e_n,x,y,\vec{q}_n)\Big\}\nn\\
&& \int d^2\vec{q}_{\bar n}\mathcal{J}^{\perp}_{\bar n}(e_{\bar n},\vec{q}_{\bar n})\delta^2(\vec{q}_T+ \vec{q}_{JS}+\vec{q}_{GS}+\vec{q}_{n}+\vec{q}_{\bar n})\nn
\eea

While most of the functions remain unchanged compared to their vacuum counterparts, we have two new/modified function $\bar{S}_G^{AB}$ and $J_n^{AB}$ defined as 
\bea
\label{ModFn}
&&\bar{S}_G^{AB}( q^{\perp}_{GS},\{x_{\perp},x^-\},\{y_{\perp},y^-\})= \langle X_S|\{\delta^2(q^{\perp}_{GS}-\mathcal{P}_{\perp})\frac{1}{\mathcal{P}_{\perp}^2}O_S^A(\hat x)\}\rho_B \frac{1}{\mathcal{P}_{\perp}^2}O_S^B(\hat y)|X_S\rangle \nn\\
&&J_n^{AB}(e_n,x,y)= \langle X_n|T\Big\{\bar \chi_n(0)\frac{\slashed{\bar n}}{2}O_n^A(x)\Big\}|0\rangle \langle 0|\bar T\Big \{O_n^B(y)\Big[\delta^2(\mathcal{P}_{\perp})\chi_n(0)\Big]\Big\}\delta^2(Q-\mathcal{P}^-)\mathcal{M}_n|X_n\rangle \nn\\
\eea
where for the rest of the paper we use  $\hat x \equiv( x^-, 0, \vec{x}_{\perp})$ and likewise $\hat y$, and
\bea
\mathcal{M}_n=\Theta(e_n- \mathcal{E}_{n \in \text{gr}})\delta^2(\vec{q}_n-\mathcal{P}_{\perp \not\in \text{gr}}) 
\eea
To derive the factorizatin formula, we have used the idea that the Wilson coefficient $C_G$ for all the Glauber operators is identical,
\bea
C_G =  8\pi \alpha_s
\eea
and
\bea
O_n^A = \sum_i O^A_{n,i} , \ \ O_S^B = \sum_j O^{B}_{S,j}
\eea

Using the translational invariance of the QGP medium we can write 
\bea
 \bar{S}_G^{AB}( q^{\perp}_{GS},\hat x,\hat y)= \int \frac{d^4k}{(2\pi)^4 k_{\perp}^4 }e^{i(\hat x-\hat y) \cdot k} D_>^{AB}(k) \delta^2(q^{\perp}_{GS}-\vec{k}_{\perp})
\eea
where $D^{AB}_{>}(k)$ is the Soft correlator in the medium
\bea
\label{SoftCor}
D_>^{AB}(k) = \int d^4x e^{-i k \cdot x} \langle X_S|O_S^{A}(x)\rho O_S^B(0)|X_S\rangle 
\eea
 We note here that this function is independent of $k^+$. \\
We can redefine $\vec{q}_n \rightarrow \vec{q}_n -\vec{k}_{\perp}$, and write a compact formula 
\bea 
\label{SigmA}
\Sigma^{(2)}_R(\vec{q}_T,e_n,e{\bar n}) &=& V\times |C_G|^2 H(Q,\mu) S(\vec{q}_T)\otimes_{q_T}\mathcal{J}^{\perp}_{\bar n}(e_{\bar n}, \vec{q}_T)\nn\\
&& \otimes_{q_T}\int \frac{d^2 k_{\perp}dk^-}{k_{\perp}^4}D^{AB}(k_{\perp},k^-) \mathcal{J}_n^{AB}(e_n, \vec{q}_T, \vec{k}_{\perp})  
\eea

where we have 
\bea
\mathcal{J}_n^{AB}(e_n, \vec{q}_T, \vec{k}_{\perp}) =  \int dk^+\int d^4x \int d^4ye^{i(\hat x-\hat y) \cdot k} \int d^2\vec{q}_n\Big\{J_n^{AB}(e_n,x,y,\vec{q}_n -\vec{k}_{\perp})\Big\}
\eea
The convolution in $\vec{q}_T $ turns into a product in impact parameter space. We can now divide and multiply by the vacuum jet function $\mathcal{J}^{\perp}_n(e_n, \vec{b})$ which allows us to factor out the vacuum cross section and define a medium structure function and a medium induced jet function
\bea
 \Sigma^{(2)}_R(\vec{q}_T,e_n,e_{\bar n})=  |C_G|^2 \int d^2\vec{b} e^{i \vec{b} \cdot \vec{q}_T}\Sigma^{(0)}(b,e_n,e_{\bar n})\int d^2 k_{\perp}S_G^{AB}(k_{\perp}) J_{n,M}^{AB}(e_n, \vec{b}, \vec{k}_{\perp})
\eea
where 
\bea
  J_{n,M}^{AB}(e_n, \vec{b}, \vec{k}_{\perp}) = \frac{1}{k_{\perp}^2}\frac{\mathcal{J}_n^{AB}(e_n, \vec{b}, \vec{k}_{\perp})}{\mathcal{J}^{\perp}_n(e_n, \vec{b})}\nn\\
	S_G^{AB}(k_{\perp}) = \int dk^-\frac{1}{k_{\perp}^2}D^{AB}(k_{\perp},k^-)
	\label{Fact_Def}
\eea


\item{{\bf Glauber insertion on the same side of the cut}}
\\\\
We can now look at the piece we get by doing two Glauber insertions on the same side of the cut starting from Eq. \ref{Sigm}. 
Here, I will consider inserting on the bra side and deal with the ket in the same manner later on.
\small
\bea
\sigma_{V}^{(2)}(t) &=& \frac{(-i)^2}{2}I^{\mu\nu}|C(Q)|^2\int d\tilde x e^{-iH^{SCET}t}\mathcal{T}\Big\{\int_0^t dt'H_{G,I_{SC}}(t')\int_0^t d\hat t H_{G,I_{SC}}(\hat t)J^{\mu}_{SCET,I_{SC}}(x_1)\Big\}|0\rangle \nn\\
&& \langle 0|\mathcal{\bar T}\Big\{J^{\nu}_{SCET,I_{SC}}(x_2)\Big\}e^{iH^{SCET}t}\nn
\eea 
\normalsize

 Following the same series of steps as in the previous section, we arrive at a similar factorization formula
\bea
\label{SigmB}
 \Sigma^{(2)}_V &=& -\frac{1}{2}V\times |C_G(Q)|^2H(Q,\mu) \int d^2\vec{q}_{JS} S(\vec{q}_{JS})\int d^2\vec{q}_{\bar n}\mathcal{J}^{\perp}_{\bar n}(e_{\bar n},\vec{q}_{\bar n})\nn\\
&&\int d^4 x\int d^4 y\bar{S}_G^{AB}(\hat x,\hat y)\int d^2 \vec{q}_nJ_n^{AB}(e_n, \vec{q}_n, x,y)\delta^2(\vec{q}_T+ \vec{q}_{JS}+\vec{q}_{n}+\vec{q}_{\bar n})
\eea
with the following definitions 
\bea
&&\bar{S}_G^{AB}(\hat x,\hat y)=  \langle X_S|\mathcal{T}\Big\{O_S^A(\hat x) O_S^B(\hat y)\Big\}\rho_B|X_S\rangle \nn\\
&&J_n^{AB}(e_n, \vec{q}_n, x,y)= \langle X_n|\mathcal{T}\Big\{\bar \chi_n(0)\frac{\slashed{\bar n}}{2}O_n^A(x)O_n^B(y)\Big\}|0\rangle \langle 0|\Big[\delta^2(\mathcal{P}_{\perp})\chi_n(0)\Big]\delta^2(Q-\mathcal{P}^-)\mathcal{M}_n|X_n\rangle \nn\\
\eea
\normalsize 
The term with Glauber insertions on the other side of the cut can now be trivially obtained from this result.
\end{itemize}


\section{Decoherence of the hard interaction from the medium}
\label{sec:Deco}

One aspect of this factorization which is different compared to a vacuum factorization result is the presence of the environment. Since the QGP medium in $not$ coherently connected with the hard interaction that produces the jet, the phase space of the jet allows the partons to go on-shell before they interact with the medium. In fact, the most dominant contribution to the cross section comes from this region of phase space. To see this explicitly we can look at the tree level result for our modified jet function defined in Eq.\ref{ModFn} that appears in the the factorized formula for $\Sigma_R^{(2)}$ in Eq.\ref{SigmA}
\small
\bea
\mathcal{J}_n^{AB(0)}&=& \int dk^+\int d^4x \int d^4y e^{-i k\cdot (\hat x- \hat y)}e^{ip\cdot(x-y)}\int \tilde d p\text{Tr}\Big[\bar u(p) T^A\frac{\slashed{\bar n}}{2}D(x)\frac{\slashed{\bar n}}{2}D^{\dagger}(y)\frac{\slashed{\bar n}}{2}T^Bu(p) \Big]\nn\\
&&\delta^2(p_{\perp})\delta(Q-p^-)\delta^2(\vec{q}_n-\vec{k}_{\perp})\nn\\
&=& \int dk^+\int d^4x \int d^4y e^{-i k\cdot (\hat x- \hat y)}e^{ip\cdot(x-y)}\int \tilde d p\nn\\
&&\text{Tr}\Big[\bar u(p)T^A \frac{\slashed{\bar n}}{2}\int d^4q \frac{\slashed n}{2}\frac{q^-e^{-iq\cdot x}}{q^2-i\epsilon}\frac{\slashed{\bar n}}{2}\int d^4 q' \frac{\slashed n}{2}\frac{(q')^-e^{iq'\cdot y}}{(q')^2+i\epsilon}\frac{\slashed{\bar n}}{2}T^Bu(p) \Big]\delta^2(p_{\perp})\delta(Q-p^-)\delta^2(\vec{q}_n-\vec{k}_{\perp})\nn\\
\eea
\normalsize
where henceforth we will reserve the notation $\tilde d p$ to mean the integral over the phase space of the on-shell massless particle p 
\bea
\tilde d p = \frac{d^3p}{2E_p} \equiv d^4p \delta^+(p^2)
\eea
The integrals over the co-ordinates x and y now sets $q=q' = p+k$. Since we are integrating over k, we see that there is a pinch singularity in the integral over $k^+$ and the dominant contribution comes from the region when the intermediate propagators(in $q=q'$) go on-shell. So we can replace the propagators with their on-shell (cut) forms, ignoring for the rest of this paper any contribution from the residual principal value. The pinch singularity seems to yield a divergence. However, we notice that the integrals over $dx^0, dy^0$ are actually over a finite time interval $\in{0,t}$ and hence, what naively appears as an infinite result actually yields a factor of t, which is just the time of propagation of the jet in the medium.\\
We can therefore revisit our factorized functions and simplify them by looking at this dominant contribution where the intermediate partons produced in the hard interaction go on-shell before interacting with the medium one at a time. This will capture all the corrections enhanced by a factor of t, which will be the dominant correction as long as $t >>1/T$, T being the energy scale of the medium partons.

\begin{itemize}
\item{$\Sigma_R^{(2)}$}\\
The medium induced jet function is defined in Eq.\ref{Fact_Def} as 
\bea
J_{n,M}^{AB}(e_n, \vec{b}, \vec{k}_{\perp}) = \frac{1}{k_{\perp}^2}\frac{\mathcal{J}_n^{AB}(e_n, \vec{b}, \vec{k}_{\perp})}{\mathcal{J}^{\perp}_n(e_n, \vec{b})}
\eea  

Since we are working in a regime where the intermediate partons are going on-shell, we can do a a further factorization of the jet function to separate out the Glauber and vacuum terms explicitly. At the same time, in anticipation of the fact that the jet function will give a result proportional to t, we can define 
\bea
\label{JetR}
&&\mathcal{J}_{n,M}^{AB}(e_n,k_{\perp},b)=\frac{1}{k_{\perp}^2 t J_n(e_n, \vec{b})}\int \frac{d^2q_{n}e^{i\vec{q}_n \cdot \vec{b}}}{(2\pi)^2}\int dk^+\int d^4x \int d^4y e^{-ik\cdot(\hat x-\hat y)}\nn\\
&\times&\langle X_n|O_n^A(x)\delta^2(\mathcal{P}^{\perp})\delta(Q-p^-)\chi_n(0)\bar \chi \frac{\slashed{\bar n}}{2}|0\rangle \langle 0|\chi_n(0)O_n^B(y)\mathcal{M}|X_n\rangle
\eea
where 
\bea
\mathcal{M} =\Theta(e_n-\mathcal{E}_{\in \text{gr}})\delta^2(\vec{q}_n-\mathcal{P}^{\perp}_{\not\in \text{gr}}-\vec{k}_{\perp}) 
\eea

and we have dropped the time ordering between the SCET and Glauber currents.
We look at the numerator(sans the Fourier transform) and we can insert a complete set of on-shell states separating the hard and Glauber operator, thus placing the intermediate partons on-shell.
\bea
&&\mathcal{J}^{AB}(e_n,k_{\perp},\vec{q}_n)=\int dk^+\int d^4x \int d^4y e^{-ik\cdot(\hat x-\hat y)}\langle X_n|O_n^A(x)|Y_n\rangle\nn\\
 &&\langle Y_n|\delta^2(\mathcal{P}_{\perp})\delta(Q-p^-)\chi_n(0)\bar \chi \frac{\slashed{\bar n}}{2}|0\rangle \langle 0|\chi_n(0)|\tilde Y_n\rangle\langle \tilde Y_n|O_n^B(y)\mathcal{M}|X_n\rangle
\eea

where its understood that there is an integral over the phase space of all the inserted states $Y_n$, $\tilde Y_n$ and the measurement $\mathcal{M}$ acts on all final state $X_n$. We can now perform the co-ordinate integrals to give
\bea
 &&\mathcal{J}^{AB}(e_n,k_{\perp}, \vec{q}_n)= \delta(p_X^--p_Y^-)\delta(p_{\tilde Y}^+-p_{Y}^+)\delta^2(k_{\perp}+p^{\perp}_Y-p^{\perp}_X)\delta(p_X^--p_{\tilde Y}^-)\delta^2(k_{\perp}+p^{\perp}_{\tilde Y}-p^{\perp}_X)\nn\\
&& \langle X_n|O_n^A(0)|Y_n\rangle \langle \tilde Y_n|O_n^B(0)\mathcal{M}|X_n\rangle \langle Y_n|\delta^2(\mathcal{P}_{\perp})\delta(Q-p^-)\bar\chi_n(0)\frac{\slashed{\bar n}}{2}|0\rangle \langle 0|\chi_n(0)|\tilde Y_n\rangle\mathcal{M}_{X_n}
\eea
The $\delta$ functions then imply $p_Y= p_{\tilde Y}$, where $p_Y = \sum_i p_{Yi}$ is the sum of momentum of the $Y_n$ states and so on.\\ 
The terms proportional to t are obtained from those diagrams where one of the partons created in the hard vertex interacts with the medium. We can therefore write $|Y_n\rangle \equiv |Y_1\rangle |\mathcal{Y}_n\rangle $ where we have explicitly separated out the parton with momentum $p_{Y_1}$ which will interact with the medium. Likewise $|\tilde Y_n\rangle \equiv |\tilde Y_1\rangle|  \tilde{\mathcal{Y}}_n\rangle $. The partons with total momenta $p_{\mathcal{Y}}$ ,$ p_{\tilde {\mathcal{Y}}_n}$ do not interact with the medium. At the same time, we need to drop any interference diagrams between the hard vertex and Glauber vertex to retain the t enhanced terms which allows us to write
\bea
 &&\mathcal{J}^{AB}(e_n,k_{\perp}, \vec{q}_n)= \delta(p_X^--p_{Y1}^--p_{\mathcal{Y}}^-)\delta(p_{\tilde{Y}_1}^++p_{\tilde{\mathcal{Y}}}^+-p_{Y_1}^+-p_{\mathcal{Y}}^+)\delta^2(k_{\perp}+p^{\perp}_{Y_1}+p^{\perp}_{\mathcal{Y}}-p^{\perp}_X)\nn\\
&& \delta(p_X^--p_{\tilde{Y}_1}^--p_{\tilde{\mathcal{Y}}}^-)\delta^2(k_{\perp}+p^{\perp}_{\tilde Y1}+p^{\perp}_{\tilde{\mathcal{Y}}}-p^{\perp}_X)\langle \tilde X_n|O_n^A(0)|Y_1\rangle \langle \tilde{Y}_1 |O_n^B(0)\mathcal{M}|\tilde X_n\rangle \nn\\
&\times& \langle Y_1 \mathcal{Y}|\delta^2(\mathcal{P}^{\perp})\delta(Q-p^-)\bar\chi_n(0)\frac{\slashed{\bar n}}{2}|0\rangle \langle 0|\chi_n(0)|\tilde{Y}_1 \tilde{\mathcal{Y}}\rangle\mathcal{M}_{X_n}\langle \bar X_n | \mathcal{Y}\rangle  \langle \tilde{\mathcal{Y}} | \bar X_n \rangle 
\eea
where we have split $|X_n \rangle  \equiv |\bar{X}_n \rangle |\tilde{X}_n\rangle $
We can now eliminate the phase space integrals over all the partons that make up $p_{\mathcal{Y}} $ and $p_{\tilde{\mathcal{Y}}}$ 
\bea
  &&\mathcal{J}^{AB}(e_n,k_{\perp}, \vec{q}_n)= \delta(p_{\tilde X}^--p_{Y_1}^-)\delta(p_{\tilde{Y}_1}^+-p_{Y_1}^+)\delta^2(k_{\perp}+p^{\perp}_{Y_1}-p^{\perp}_{\tilde{X}})\delta(p_{\tilde{X}}^--p_{\tilde{Y}_1}^-) \delta^2(k_{\perp}+p^{\perp}_{\tilde Y1}-p^{\perp}_{\tilde X})\nn\\
&&\langle \tilde X_n|O_n^A(0)|Y_1\rangle \langle \tilde{Y}_1 |O_n^B(0)\mathcal{M}|\tilde X_n\rangle \langle Y_1 \bar{X}_n|\delta^2(\mathcal{P}^{\perp})\delta(Q-p^-)\bar\chi_n(0)\frac{\slashed{\bar n}}{2}|0\rangle \langle 0|\chi_n(0)|\tilde{Y}_1\bar{X}_n\rangle \mathcal{M}_{X_n}\nn\\
\eea
which tells us that $p_{Y1} =p_{\tilde Y1} =p_X-p_{\bar X}-k =p_{\tilde X}-k$. We can now perform the phase space integrals over $p_{Y1}$, $p_{\tilde Y1}$, which yields a redundant $\delta$ function which is just the factor of t that we expected. We can therefore write 
\bea
  &&\mathcal{J}^{AB}(e_n,k_{\perp}, \vec{q}_n)=\frac{t}{[p_{\tilde X}^-]^2}\langle p_{\tilde X}-k, \bar{X}_n|\delta^2(\mathcal{P}_{\perp})\delta(Q-p^-)\bar\chi_n(0)\frac{\slashed{\bar n}}{2}|0\rangle \langle 0|\chi_n(0)|p_{\tilde X}-k, \bar{X}_n\rangle  \nn\\
	&\times&  \langle \tilde X_n|O_n^A(0)|p_{\tilde X}-k\rangle \langle p_{\tilde X}-k|O_n^B(0)|\tilde X_n\rangle \mathcal{M}_{X_n}
\eea

which allows us to write our factorization formula for $\Sigma^{(2)}_R$ as 
\bea
 \Sigma^{(2)}_R(\vec{q}_T,e_n,e_{\bar n})=  t|C_G|^2 \int d^2b e^{i\vec{q}_T\cdot \vec{b}}\Sigma^{(0)}(\vec{b},e_n,e_{\bar n})\int d^2 k_{\perp}S_G^{AB}(k_{\perp}) \mathcal{J}_{n,R}^{AB}(e_n, b, \vec{k}_{\perp})
\eea
with 
\bea
\label{Jetr}
 \mathcal{J}_{n,R}^{AB}(e_n, \vec{q}_T, \vec{k}_{\perp})&=&\frac{1}{\mathcal{J}^{\perp}_n(e_n,b)k_{\perp}^2}\frac{1}{[p_{\tilde X}^-]^2}\int \frac{d^2q_ne^{-i\vec{q}_n\cdot \vec{b}}}{(2\pi)^2}\langle p_{\tilde X}-k, \bar{X}_n|\delta^2(\mathcal{P}_{\perp})\delta(Q-p^-)\bar\chi_n(0)\frac{\slashed{\bar n}}{2}|0\rangle \nn\\
	&\times&  \langle 0|\chi_n(0)|p_{\tilde X}-k, \bar{X}_n\rangle  \langle \tilde X_n|O_n^A(0)|p_{\tilde X}-k\rangle \langle p_{\tilde X}-k|O_n^B(0)|\tilde X_n\rangle \mathcal{M}_{X_n}
\eea


 \item{$\Sigma_V^{(2)}$}\\
The factorized form for this piece was derived in Eq. \ref{SigmB} in terms of the functions $S_G$ convoluted with a a jet function with the following definitions 
\bea
&&\bar{S}_G^{AB}(\hat x,\hat y)=  \langle X_S|T\Big\{\frac{1}{\mathcal{P}_{\perp}^2}O_S^A(\hat x) \frac{1}{\mathcal{P}_{\perp}^2}O_S^B(\hat y)\Big\}\rho_B|X_S\rangle \nn\\
&&J_n^{AB}(e_n, \vec{q}_n, x,y)= \langle X_n|\mathcal{T}\Big\{\chi_n(0)\frac{\slashed{\bar n}}{2}O_n^A(x)O_n^B(y)\Big\}|0\rangle \langle 0|\bar{\mathcal{T}}\Big \{\bar \chi_n(0)\Big\}\delta^2(\mathcal{P}_{\perp})\delta^2(Q-\mathcal{P}^-)\mathcal{M}_V|X_n\rangle \nn\\
\eea
where 
\bea
\mathcal{M}_V&=&\Theta(e_n- \mathcal{E}_{\in n,\text{gr}})\delta^2(\vec{q}_n-\mathcal{P}^{\perp}_{\not\in n,\text{gr}}) 
\eea
Once again we are going to put the intermediate particles on-shell before they interact with the medium. This allows us to simplify our jet function by inserting a complete basis on on-shell states
\bea
 J_n^{AB}(x,y)&=& \langle X_n|\mathcal{T}\Big\{O_n^A(x)O_n^B(y)\Big\}|Y_n\rangle \langle Y_n|\chi_n(0)\frac{\slashed{\bar n}}{2}|0\rangle \langle 0|\bar \chi_n(0)\delta^2(\mathcal{P}_{\perp})\delta^2(Q-\mathcal{P}^-)\mathcal{M}_V|X_n\rangle \nn\\
\eea
Following the same logic as for $\Sigma_R^{(2)}$, we look at the scenario where a single parton created in the hard vertex($Y_1$) interacts with the medium at a time ignoring any interference diagrams between the two vertices. 
\bea
 J_n^{AB}(x,y)&=& \langle \tilde{X}_n|\mathcal{T}\Big\{O_n^A(x)O_n^B(y)\Big\}|Y_1\rangle \langle Y_1 \mathcal{Y}|\chi_n(0)\frac{\slashed{\bar n}}{2}|0\rangle  \nn\\
&\times& \langle 0|\bar \chi_n(0)\delta^2(\mathcal{P}_{\perp})\delta^2(Q-\mathcal{P}^-)\mathcal{M}_V|\tilde X_n \bar{X}_n\rangle\langle \bar X_n|\mathcal{Y}\rangle 
\eea
which allows us to eliminate the phase space integrals over $\mathcal{Y}$.  Once again we have split $|X_n \rangle \equiv |\bar X_n\rangle \tilde X_n \rangle $. At the same time since we are ignoring interference diagrams between the hard and Glauber vertices, this necessarily means that $\tilde X_n$ is a single particle state. 
\bea
 J_n^{AB}(x,y)&=& \langle \tilde{X}_n|\mathcal{T}\Big\{O_n^A(x)O_n^B(y)\Big\}|Y_1 \rangle \langle Y_1 \bar {X}_{n}|\chi_n(0)\frac{\slashed{\bar n}}{2}|0\rangle \nn\\
&&\langle 0|\bar \chi_n(0)\delta^2(\mathcal{P}_{\perp})\delta^2(Q-\mathcal{P}^-)\mathcal{M}_V|\tilde X_n \bar{X}_n\rangle 
\eea
We can now  look at the combination of the Soft and jet function and completely separate out the medium dependent piece 
\bea
\mathcal{G}&=&\int d^4 x\int d^4 y  S_G^{AB}(\hat x,\hat y) J_n^{AB}(x,y)\nn\\
&=&2 \int d^4 x\int d^4 y \Theta(x^0-y^0)\langle X_S|\frac{1}{\mathcal{P}_{\perp}^2}O_S^A(\hat x) \frac{1}{\mathcal{P}_{\perp}^2}O_S^B(\hat y)\rho_B|X_S\rangle \langle \tilde{X}_n|O_n^A(x)O_n^B(y)|Y_1 \rangle \nn\\
&&\langle Y_1 \bar {X}_{n}|\chi_n(0)\frac{\slashed{\bar n}}{2}|0\rangle \langle 0|\bar \chi_n(0)\delta^2(\mathcal{P}_{\perp})\delta^2(Q-\mathcal{P}^-)\mathcal{M}_V|\tilde X_n \bar{X}_n\rangle
\eea
We can now write $\Theta(t)= 1/2(1+\text{sgn}(t))$ separating out the terms corresponding to Unitary evolution from the dissipative. 
\bea
\mathcal{G} = \mathcal{G}_U+ \mathcal{G}_D 
\eea
The piece corresponding to unitary evolution reads 
\bea
  \mathcal{G}_U&=&\int d^4 x\int d^4 y \text{sgn}(x^0-y^0)\langle X_S|\frac{1}{\mathcal{P}_{\perp}^2}O_S^A(\hat x) \frac{1}{\mathcal{P}_{\perp}^2}O_S^B(\hat y)\rho_B|X_S\rangle\langle \tilde{X}_n|O_n^A(x)O_n^B(y)|Y_1 \rangle \nn\\
&&\langle Y_1 \bar {X}_{n}|\chi_n(0)\frac{\slashed{\bar n}}{2}|0\rangle \langle 0|\bar \chi_n(0)\delta^2(\mathcal{P}_{\perp})\delta^2(Q-\mathcal{P}^-)\mathcal{M}_V|\tilde X_n \bar{X}_n\rangle
 \eea
This piece cancels out with the corresponding Unitary evolution term from double Glauber insertions on the other side of the cut (By interchanging the variables $p_{Y_1}$ and $p_{\tilde X_n}$). The Dissipative piece, on the other hand no longer contains any time ordering. Then, as before in Eq.\ref{SoftCor}, using the translational invariance of the QGP medium,
 \bea
 \langle X_S|\frac{1}{\mathcal{P}_{\perp}^2}O_S^A(x_{\perp},x^-)\frac{1}{\mathcal{P}_{\perp}^2} O_S^B(y_{\perp},y^-)\rho_B|X_S\rangle = \int \frac{d^4k}{(2\pi)^4 k_{\perp}^4 }e^{i(\hat x-\hat y) \cdot k} D_>^{AB}(k)\nn
\eea

 The factorization formula now becomes 
\bea
\Sigma^{(2)}_V &=& -\frac{1}{2}V\times |C_G(Q)|^2H(Q,\mu) \int d^2\vec{q}_{JS} S(\vec{q}_{JS}) \int d^2\vec{q}_{\bar n}\mathcal{J}^{\perp}_{\bar n}(e_{\bar n},\vec{q}_{\bar n})\nn\\
&&\int \frac{d^4k}{(2\pi)^4 k_{\perp}^4 } D_>^{AB}(k)\int d^2 \vec{q}_n\int d^4 x\int d^4 ye^{i(\hat x-\hat y) \cdot k}J_n^{AB}(e_n,\vec{q}_n,x,y)\nn\\
&\times& \delta^2(\vec{q}_T+ \vec{q}_{JS}+\vec{q}_{n}+\vec{q}_{\bar n})
\eea

with 
\bea
J_n^{AB}(e_n,\vec{q}_n,x,y) = \langle \tilde{X}_n|O_n^A(x)O_n^B(y)|Y_1\rangle \langle Y_1 \bar {X}_{n}|\chi_n(0)\frac{\slashed{\bar n}}{2}|0\rangle \langle 0|\bar \chi_n(0)\delta^2(\mathcal{P}_{\perp})\delta^2(Q-\mathcal{P}^-)\mathcal{M}_V|\tilde X_n \bar{X}_n\rangle \nn
\eea
We can now insert a complete set so states and perform the integrals over $x,y$, This will enable us to eliminate the phase space integrals $p_{Y1}$ and $\tilde X_n$, resulting in a factor of t.
\bea
&&\int d^4x d^4 y e^{-i k\cdot (\hat x- \hat y )}J_n^{AB}(e_n, \vec{q}_n x,y) \nn\\
&=& \frac{t}{[p_{\tilde Y}^-]^2} \langle p_{\tilde Y}+k|O_n^A(0)|\tilde Y_n \rangle \langle \tilde Y_n|O_n^B(0)|p_{\tilde Y}+k \rangle \langle p_{\tilde Y}+k,\bar {X}_{n}|\chi_n(0)\frac{\slashed{\bar n}}{2}|0\rangle\nn\\
&&\langle 0|\bar \chi_n(0)\delta^2(\mathcal{P}_{\perp})\delta^2(Q-\mathcal{P}^-)\mathcal{M}_V|p_{\tilde Y}+k,\bar{X}_n\rangle
\eea

We can factor out the vacuum cross section as before and write,
\bea
 \Sigma^{(2)}_V &=& -\frac{1}{2} t\times |C_G(Q)|^2 \int d^2be^{i\vec{b} \cdot \vec{q}_T}\Sigma^{(0)}(\vec{b},e_n,e_{\bar n})\int \frac{d^2k_{\perp}}{(2\pi)^4 } S_G^{AB}(k_{\perp})\mathcal{J}^{AB}_{n,V}(e_n, b,\vec{k}_{\perp})\nn\\
\eea
where the Soft function $S_G^{AB}$ is identical to the one for $\Sigma_R^{(2)}$ defined in Eq.\ref{Fact_Def} and 
\bea
\label{JetV}
\mathcal{J}^{AB}_{n,V}(b) &= & \frac{1}{k_{\perp}^2 \mathcal{J}^{\perp}_n(e_n,b)} \frac{1}{[p_{\tilde Y}^-]^2}\int \frac{d^2q_ne^{-i\vec{q}_n\cdot \vec{b}}}{(2\pi)^2} \langle p_{\tilde Y}+k|O_n^A(0)|\tilde Y_n \rangle \langle \tilde Y_n|O_n^B(0)|p_{\tilde Y}+k \rangle \nn\\
&&\langle p_{\tilde Y}+k,\bar {X}_{n}|\chi_n(0)\frac{\slashed{\bar n}}{2}|0\rangle \langle 0|\bar \chi_n(0)\delta^2(\mathcal{P}_{\perp})\delta^2(Q-\mathcal{P}^-)\mathcal{M}_V|p_{\tilde Y}+k,\bar{X}_n\rangle
\eea

\end{itemize}
We can now combine the two simplified terms to write the final form of our factorization formula 
\bea
\Sigma_R^{(2)}+ \Big\{\Sigma_V^{(2)}+ c.c\Big\} = t\times |C_G(Q)|^2 \Sigma^{(0)}(\vec{q}_T,e_n,e_{\bar n})\otimes_{q_T}\int d^2k_{\perp} S_G^{AB}(k_{\perp})\mathcal{J}^{AB}_{n,\text{Med}}(e_n, \vec{q}_T,k_{\perp})\nn\\
\eea

with an effective medium induced jet function 
\bea
\label{JetMed}
 \mathcal{J}^{AB}_{n,\text{Med}}(e_n, \vec{q}_T) = \mathcal{J}^{AB}_{n,R}(e_n, \vec{q}_T,\vec{k}_{\perp})-\mathcal{J}^{AB}_{n,V}(e_n, \vec{q}_T,\vec{k}_{\perp})
\eea
in terms of $ \mathcal{J}^{AB}_{n,R}$ and $ \mathcal{J}^{AB}_{n,V}$ defined in Eqn. \ref{Jetr} and Eq. \ref{JetV}.

In the next section, we will consider the one loop corrections for the functions in this factorization formula.


\section{One loop results for medium induced functions}
\label{sec:JetMod}
The factorization formula at next to leading order in the Glauber operator expansion reveals two new functions beyond those already present in the vacuum result:  A medium Soft function and a medium induced jet function. Of these, the medium soft function $S_G^{AB}$ defined in Eq. \ref{Fact_Def} contains all the information about the structure of the medium and is independent of the observable defined on the final state jet. Hence, it can be thought of as a universal structure function for the medium. An explicit computation of this function now requires us to assume some form for the medium density matrix. A relevant choice is a medium in thermal equilibrium at temperature $T<<Q$ where the EFT applies for which 
\bea
\rho_B = \frac{e^{-\beta H_S}}{\text{Tr}\Big[e^{-\beta H_S}\Big]}
\eea
where $H_S$ is the Soft sector Hamiltonian which is identical to the full QCD Hamiltonian.

The medium induced jet function encodes the modification of the jet due to the interaction with the medium. The complete set of corrections for this function at one loop involves: 
\begin{itemize}
\item{} Elastic collisions of the jet partons with the medium.
\item{} Medium induced radiation. 
\end{itemize}
In this paper, we will only consider the one loop corrections for the hard vertex, which correspond to the corrections from elastic collisions  of the vacuum evolved jet partons with the medium. The corrections to the Glauber vertex which corresponds to medium induced radiation will be presented elsewhere \cite{VarB}. Since the medium Soft function is a correlator of the Galuber Soft operator, we therefore compute the Soft function to tree level. With the assumption of a thermal medium, the tree level soft function is then simply the Wightman function in the thermal bath. This was already computed in \cite{Vaidya:2020cyi}. We present the result here for convenience. 
\bea
S_G^{AB,(0)}(\vec{k}_{\perp})&=& \frac{\delta^{AB}}{2k_{\perp}^2} \int \frac{d^2p_{\perp}}{(2\pi)^3}\int_0^{\infty} dp^+ \Bigg\{n_F\left(|\frac{p^+}{2}+\frac{p_{\perp}^2}{2}|\right)\Big[1-n_F(|\frac{p^+}{2}+\frac{(\vec{p}_{\perp}+\vec{k}_{\perp})^2}{2p^+}|)\Big]\nn\\
&+&n_F(|\frac{p^+}{2}+\frac{(\vec{p}_{\perp}+\vec{k}_{\perp})^2}{2p^+}|)\Big[1-n_F\left(|\frac{p^+}{2}+\frac{p_{\perp}^2}{2}|\right)\Big]\Bigg\}\nn
\eea
This result is written in terms of the Fermi distribution function assuming that the thermal bath has fermions. A complete calculation would also include a similar term with Bose distribution functions to account for gluons in the thermal bath, but for simplicity we will not consider those terms till we do a phenomenological analysis.\\
We can now look at the medium jet function, specifically considering the one loop corrections from elastic scattering.
At tree level we have in impact parameter space 
\bea
\mathcal{J}_{n,\text{Med}}^{AB,(0)}(\vec{k}_{\perp},b,e_n) = \frac{e^{-i\vec{k}_{\perp}\cdot \vec{b}}-1}{k_{\perp}^2} 
\eea

The details of the one loop computation for the elastic collisions are computed in Appendix \ref{App:JetM}.
\bea
\mathcal{J}_{n,\text{Med}(1)}^{AB}&=& \frac{\frac{1}{2}\delta^{AB}}{k_{\perp}^2}\left(Q(b) + G(b)\right)
\eea
where the result is written in terms of Quark and Gluon Glauber operator contributions 
\small
\bea
 Q(b)&=&\frac{\alpha_s C_F e^{-i\vec{k}_{\perp}\cdot \vec{b}}}{2 \pi} \Bigg[\int_{z_c}^{1-z_c}dz p_{gq}(z)\Big\{\Theta(z_c-\frac{y}{e_n+y})\ln \left(\frac{-B(z)}{e_n(1-z)}\right)-\Theta(z_c-\frac{e_n}{e_n+y})\ln \frac{B(z)e_n}{yM}\Big\}\nn\\
&+&\Theta( \frac{e_n}{e_n+y}-z_c)\Theta(\frac{y}{e_n+y}-z_c)\Big\{\int_{z_c}^{\frac{e_n}{e_n+y}}dz p_{gq}(z)\ln \left(\frac{-B(z)}{e_n(1-z)}\right)-\int_{\frac{e_n}{e_n+y}}^{1-z_c}dz p_{gq}(z)\ln \frac{B(z)e_n}{yM}\Big\}\Bigg]\nn\\
&+& \frac{\alpha_sC_F}{2\pi}\left(e^{-i\vec{k}_{\perp}\cdot \vec{b}}-1\right)\int_{1-z_c}^{1}p_{gq}(z)\ln \frac{m_D^2b^2e^{2\gamma_E}(1-z)}{4}
\eea
The Gluon operator contribution G(b) is given by the difference $R_g-V_g$, where 
\bea 
R_g&=& \frac{\alpha_s N_c e^{-i\vec{k}_{\perp}\cdot \vec{b}}}{2 \pi} \Bigg[\int_{z_c}^{1-z_c}dz p_{gq}(z)\Big\{\Theta(z_c-\frac{y}{e_n+y})\ln\frac{A(z)}{M}-\Theta(z_c-\frac{e_n}{e_n+y})\ln \left(\frac{-A(z)}{y(1-z)}\right)\Big\}\nn\\
&-&\Theta( \frac{e_n}{e_n+y}-z_c)\Theta(\frac{y}{e_n+y}-z_c)\Big\{\int_{z_c}^{\frac{y}{e_n+y}}dz p_{gq}(z)\ln \left(\frac{-A(z)}{y(1-z)}\right)-\int_{\frac{y}{e_n+y}}^{1-z_c}dz p_{gq}(z)\ln \frac{A(z)}{M}\Big\}\Bigg]\nn\\
&-& \frac{\alpha_sN_c}{2\pi}e^{-i\vec{k}_{\perp}\cdot \vec{b}}\int_{1-z_c}^{1}p_{gq}(z)\ln \frac{m_D^2b^2e^{2\gamma_E}(1-z)}{4}
\eea 
\normalsize
written in terms of
\bea
p_{gq}(z) = \frac{1+ (1-z)^2}{z}, \ \ \  y = \frac{4k_{\perp}^2}{Q^2}, \ \ \ M =\frac{4m_D^2}{Q^2}, \ \ A(z) =e_nz-(1-z)y , \ \ B(z) =zy-e_n(1-z)\nn
\eea 
and $V_g$ is obtained by simply evaluating $R_g$ ay $k_{\perp}=0$.
The integrals over z can be done exactly analytically but are not too illuminating.
The contributions Q(b) and G(b) go to 0 as $k_{\perp}$ goes to 0 so that it is a purely medium induced result. There are two effects to note here 
\begin{itemize}
\item These radiative corrections result from purely elastic collisions of the jet with the medium and do not include any medium induced Bremsstrahlung, which will be accounted for in a companion paper \cite{VarB}. The result does not have any UV or rapidity divergences and hence does not induce any anomalous dimension for the medium jet function. Therefore, there is no resummation to be done for this function.   
\item 
The result is sensitive to the IR cut-off $m_D$ in the form of a logarithm of $\sim \ln e_n/M$, $\ln b m_D$. This appears due to the IR pole when the gluon becomes collinear to the quark. For a purely vacuum background, the collinear pole cancels between real and virtual diagrams for an IRC safe observable like the jet mass. However, incoherent elastic collision of the jet parton with the medium can transfer enough transverse momentum which can remove this pole from the real diagram if the resulting jet mass is greater than the imposed values of $e_n$. This leaves the contribution from the virtual diagram uncanceled resulting in the sensitivity to $m_D$. Similarly for the $q_T$ measurement, the collinear pole shifts away from $q_T=0$ due to the gain of additional $k_T$ from the medium resulting in non-cancellation. A possible resummation of these logarithms would require us to separate the scale $e_n, q_T$ from $M^2$, which will require a further matching to the EFT at the scale $m_D$ which is beyond the scope of this paper. 
\end{itemize}


\section{The master equation}
\label{sec:Master}

We can now gather all the pieces and write down the Master equation for the density matrix evolution. Since the radiative corrections to the medium induced corrections that we have considered in the previous section do not lead to any new anomalous dimensions, we simply include these corrections as a fixed order correction on top of a resummed vacuum cross section (Appendix \ref{App:vacuum}). Combining all the pieces from the previous section and using them in Eq. \ref{LEQ}, we can write
\bea
 &&\text{Tr}[\rho(t) \mathcal{M}]\equiv \Sigma(e_n,\vec{q}_T,t)\nn\\
&=& V\times \Bigg[\Sigma^{(0)}_{\text{Resum}}(e_n,\vec{q}_T)+ t |C_G(Q)|^2\Sigma_{\text{Resum}}^{(0)}(e_n,\vec{q}_T)\otimes_{q_T}\int d^2k_{\perp}S_G^{AB}(k_{\perp})\mathcal{J}_{n,\text{Med}}^{AB}\Bigg]+O(H_G^3)+..\nn
\eea

 We can now relate the trace over the density matrix to the scattering cross section, noting that 
\bea
\frac{d\sigma}{de_n d^2 \vec{q}_T}(t) =\mathcal{N}\frac{ \text{Tr}[\rho(t) \mathcal{M}]}{V}
\eea
where $\mathcal{N}$ is a normalization factor that depends on the initial state kinematics which we can absorb in the born level cross section. Notice here that there is still a time dependence in the cross section which is unusual but we interpret this as the time of propagation through the quark gluon plasma which in turn will depend on the length over which the jet traverses through the QGP medium. Hence, this should be treated as a length scale. 
\bea
 \frac{d\sigma(e_n,e_{\bar n},t)}{d^2\vec{q}_T}&=& \frac{d\sigma(e_n,e_{\bar n})}{d^2\vec{q}_T}\Big|^{\text{Vac}}_{\text{Resum}}+ t |C_G(Q)|^2\frac{d\sigma(e_n,e_{\bar n})}{d^2\vec{q}_T}\Big|^{\text{Vac}}_{\text{Resum}}\otimes_{q_T}\int d^2k_{\perp}S_G^{AB}(k_{\perp})\mathcal{J}_{n,\text{Med}}^{AB}\nn
\eea
This equation has an iterative structure, which can be thought of as expressing the cross section at time t in terms of the cross section at t=0. We can therefore consider this as a Markovian evolution equation for our jet over a time scale t. Taking the limit $t \rightarrow 0$, which is justified in the Markovian approximation as explained in \cite{Vaidya:2020cyi}, we can write an evolution equation for our observable as a function of the time of propagation in the QGP medium.
\bea
 \partial_t\frac{d\sigma(e_n,e_{\bar n},t)}{d^2\vec{q}_T}&=& |C_G(Q)|^2\frac{d\sigma(e_n,e_{\bar n},t)}{d^2\vec{q}_T}\otimes_{q_T}\int d^2k_{\perp}S_G^{AB}(k_{\perp})\mathcal{J}_{n,\text{Med}}^{AB}(e_n ,\vec{q}_T, \vec{k}_{\perp})
\eea

The solution for this equation will resum the leading "t" enhanced terms for all higher power contributions of $H_G$. The physical picture we have is a summation of multiple $incoherent$ interactions of the jet with the medium. This can now be solved by going to impact parameter space. This yields the result   
\bea
\frac{d\sigma(e_n,e_{\bar n}, t)}{d^2\vec{q}_T}= \int d^2 \vec{b} e^{i \vec{b}\cdot \vec{q}_T}\sigma_{\text{resum}}^{\text{vac}}(e_n, e_{\bar n},\vec{b})e^{ t|C_G(Q)|^2  \int d^2k_{\perp}S_G^{AB}(k_{\perp})\mathcal{J}_{n,\text{Med}}^{AB}(e_n ,\vec{b},\vec{k}_{\perp})}
\label{Master}
\eea
where $\sigma^{\text{vac}}(e_n, e_{\bar n},\vec{b})$ is just the inverse Fourier transform for the resummed vacuum cross section.
\bea
\sigma^{\text{vac}}_{\text{resum}}(e_n, e_{\bar n},\vec{b}) = \int \frac{d^2\vec{q}_Te^{-i \vec{q}_T \cdot \vec{b}}}{(2\pi)^2}\frac{d\sigma(e_n,e_{\bar n})}{d^2\vec{q}_T}\Big|^{\text{Vac}}_{\text{Resum}}
\eea
Eq.\ref{Master} is the main result of this paper.  While the result gives us a closed form expression, we still need to do the integrals numerically from this point onwards. A realistic comparison with data will need us to include the effects of nuclear/ hadronic pdfs and we leave a detailed phenomenological study based on this framework for the future. At the same time, the corrections to the Glauber vertex, which we have not included in this paper will likely induce new UV and rapidity divergences in our medium soft and jet functions which will also need to be resummed.


\section{Summary and Outlook}
\label{sec:conclusion}

In this paper, I develop an Effective Field Theory (EFT) framework to compute jet substructure observables for heavy ion collision experiments. 
I consider dijet events that happen at the periphery of the collision so that one of the jets evolves through vacuum while the other travels through the Quark Gluon Plasma medium that is created in the background. The jets are groomed using a grooming algorithm in order to mitigate effects of soft contamination from Multi-parton interactions as well as the QGP medium. This means that the final state measurements do not include any soft hadrons from the cooling QGP medium. We can then only follow the evolution of the reduced density matrix of the jet tracing over the QGP bath. This effectively treats the jet as an open quantum system interacting with a thermal bath and allows us to derive a Lindblad type master equation for the density matrix evolution.\\
I measure two quantities on the final state di-jet configuration:  The transverse momentum imbalance between the jets as well as a jet mass constraint imposed on each jet which restricts the radiation inside the large radius jets to a collinear core. This automatically ensures that we select events where the radiation is not sensitive to the edge of the jet. This translates to an insensitivity to jet selection bias and allows a direct comparison of jet substructure modification for the $same$ hard events in pp and HIC.

I note that the dominant contribution to the evolution of the jet in the medium comes from the regime where the partons created in the hard interaction and subsequent vacuum shower go on-shell before they interact with the medium. This happens due to the lack of quantum coherence between the hard and medium interactions. I derive a factorization formula for the observable assuming a forward scattering of the jet. This allows us to cleanly separate out the physics at different scales in terms of manifestly gauge invariant operators. The factorization holds independent of the exact form of the medium density matrix, with the only qualification that the medium be homogeneous over the length and time scales probed by a single coherent interaction of the jet with the medium.  The final formula is derived in terms of the vacuum cross section and a medium soft and jet function. The factorization formula will serve as a template for any jet substructure observable that we may wish to compute and reveals certain universal features.
\begin{itemize}
\item{} The physics of the medium is completely captured by a medium Soft function which is a correlator of Soft scaling fields in the background of the medium density matrix. This can be thought of as a "Medium Structure Function" (MSF) which is analogous to the Parton Distribution function(PDF) which encodes the longitudinal structure of hadrons. In this paper, I have only computed this to tree level in a thermal background. The radiative corrections will be considered in a companion paper \cite{VarB} which is likely to lead to both UV and rapidity anomalous dimensions for this function. 

\item{} The modification of the jet due to its interaction with the medium is captured by the medium jet function. In this paper, we have only considered the corrections from elastic collisions of the jet partons with the medium. This leads to UV finite corrections, which are however sensitive to medium IR cut-off scale $m_D$.  In a high temperature-weak coupling regime, $m_D \sim gT$ is hierarchically separated from the scale of the observable and leads to large logarithms. These can be resummed by doing a further matching of the jet function to an EFT at scale $m_D$, which we leave for the future. The other radiative corrections arise due to medium induced Bremsstrahlung will be computed in another paper \cite{VarB}.

\end{itemize}

The jet-medium interaction leads to an emergent time scale $t_I \sim 1/(\alpha_s T)$ which is the time over which the jet undergoes O(1) evolution. If the medium size is comparable to or larger than this length scale, then it becomes necessary to resum multiple incoherent interactions of the jet with the medium. This is achieved by solving a Lindblad type evolution equation which can be solved analytically in terms of a resummed vacuum cross section and medium induced functions.

While I have treated the initial hard interaction as an $e^+e^-$ collision for ease of analysis, this can be easily extended to the realistic case of hadronic collisions which will also require us to input nuclear pdfs for comparison with data. We leave the detailed phenomenological analysis based on this framework for future work. \\

\acknowledgments
I thank Xiaojun Yao and Krishna Rajagopal for useful discussions on several conceptual aspects. This work is supported by the Office of Nuclear Physics of the U.S. Department of Energy under Contract DE-SC0011090 and Department of Physics, Massachusetts Institute of Technology.


\appendix

\section{Operator definitions and one loop results for vacuum evolution}
\label{App:vacuum}

In this appendix we give the operator definitions of the factorization elements that appear in the vacuum  and their NLO expansions. From those we determine the renormalization group equations, and corresponding anomalous dimensions. Part of these results are already available in literature. The jet function is the one we compute here while quoting results for other functions.
\subsection{Hard function}
The one loop hard function for the process $e^+e^- \rightarrow q \bar q $ is (\cite{Bauer:2011uc},\cite{Ellis:2010rwa})
\bea
H= 1+ \frac{\alpha_sC_F}{2\pi}\left(-\ln^2 \frac{\mu^2}{Q^2}-3 \ln \frac{\mu^2}{Q^2} -8+\frac{7}{6}\pi^2\right)
\eea 
which yields the anomalous dimension
\bea
\gamma_{\mu}^H =  -2\frac{\alpha_sC_F}{\pi}\ln \frac{\mu^2}{Q^2} -3\frac{\alpha_sC_F}{\pi}
\eea

\subsection{Jet function}
The vacuum jet function with a cumulative jet mass measurement along with transverse momentum imbalance is given by
\bea
\mathcal{J}_n^{\perp}(\vec{q}_T, e_n) &=& \frac{(2\pi)^3}{N_c}\text{tr}\langle 0|\frac{\slashed{\bar n}}{2}\chi_n\delta(Q-\mathcal{P}^-)\delta^2(\mathcal{P}^{\perp})\delta^2(\vec{q}_T- \mathcal{P}^{\perp}_{\notin n,\text{gr}})\Theta(e_n-\mathcal{E}_{\in n,\text{gr}})|X_n\rangle \langle X_n|\bar \chi_n(0)|0\rangle  \nn
\eea
We will work in impact parameter space and give results for 
\bea
\mathcal{J}_n^{\perp}(\vec{b},e_n) = \int \frac{d^2\vec{q}_T}{(2\pi)^2}e^{-i \vec{q}_T \cdot \vec{b}} \mathcal{J}_n^{\perp}(\vec{q}_T,e_n) 
\eea
At tree level,
\bea
\mathcal{J}^{\perp{(0)}}_n &=& \frac{1}{(2\pi)^2}
\eea
\begin{figure}
     \centering
     \includegraphics[width=\linewidth]{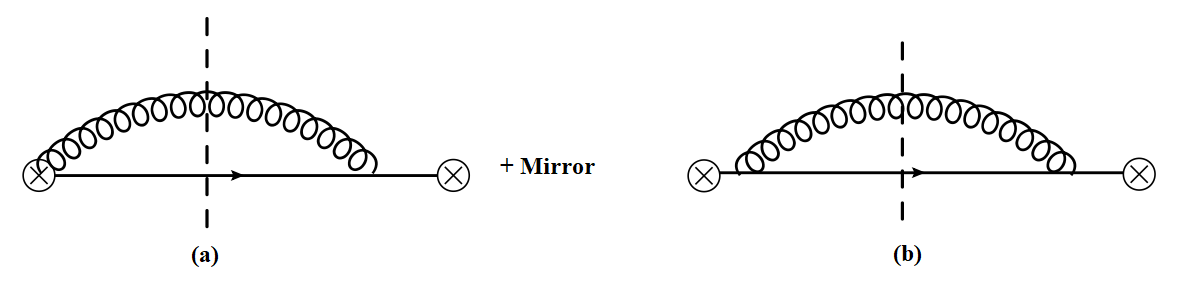}
     \caption{Real emission diagrams}
     \label{vac}
\end{figure}
At one loop we have two real and two virtual diagrams. Since we know that the vacuum cross section is an IR finite quantity, we can directly use dim reg for regulating both UV and IR divergences. In that case, the virtual diagrams are scaleless and evaluate to 0 in dim. reg.
The two real diagrams are shown in Fig.\ref{vac}.
\bea
R_a &=& 4g^2C_F\mu^{2\epsilon} \nu^{\eta}\int \frac{d^dq}{(2\pi)^{d-1}}\delta(q^2)\frac{Q|q^-|^{-\eta}}{q^-Q(q^++\frac{q_{\perp}^2}{Q-q^-})}\mathcal{M}_J
\eea
\bea
R_b&=&  2g^2C_F\mu^{2\epsilon}\int \frac{d^dq}{(2\pi)^{d-1}}\delta(q^2)\frac{q_{\perp}^2}{q^-(Q-q^-)^2(q^++\frac{q_{\perp}^2}{Q-q^-})^2}\mathcal{M}_J
\eea
\bea
\mathcal{M}_J&=&\Big\{\Theta(q^--Qz_{c})\Theta((Q-q^-)-Qz_{c})\Theta(E_J^2e_n -Q(q^++\frac{q_{\perp}^2}{Q-q^-}))\nn\\
&+&\delta^2(\vec{q}_T-\vec{q}_{\perp})\Theta((Q-q^-)-Qz_{c})\Theta(Qz_{c}-q^-)+\delta^2(\vec{q}_T+\vec{q}_{\perp})\Theta(q^--Qz_{c})\Theta(Qz_{c}-(Q-q^-))\Big\}\nn\\
\eea

This evaluates to the renormalized result in impact parameter space,
\bea
\mathcal{J}_n^{\perp(1)}(e_n,b,\mu,\nu)&=& -\frac{\alpha_sC_F}{(2\pi)^3} \ln \frac{\mu^2}{E_J^2e_n}\left( 2\ln \frac{1-z_c}{z_c}- 2(1-z_c)+2z_c +\frac{(1-z_c)^2}{2}-\frac{z_c^2}{2}\right) \nn\\
&-&\frac{\alpha_sC_F}{(2\pi)^3}\ln \frac{\mu^2b^2e^{2\gamma_E}}{4} \left(-2\ln \frac{\nu}{Qz_c}-4z_{c}-2\ln (1-z_c)+\frac{z_c^2}{2}+\frac{1}{2}-\frac{(1-z_c)^2}{2} \right)\nn\\
\eea

and therefore gives the anomalous dimensions 
\bea
\gamma_{\mu}^J&=&  \frac{\alpha_sC_F}{\pi}\left( 2\ln \frac{\nu}{Q}+ \frac{3}{2}\right)  \nn\\
\gamma_{\nu}^J&=&   \frac{\alpha_sC_F}{\pi}\ln \frac{\mu^2b^2e^{2\gamma_E}}{4} 
\eea


\subsection{Soft function}
\label{GSoft}

The  soft function that appears in the factorization theorems in eq.~(\ref{SigmZ})  is defined in eq.~(\ref{eq:soft}) and it has been calculated in several schemes at higher orders in QCD, as in~\cite{Echevarria:2015byo}.
and satisfies the following renormalization group equations 
\begin{align}
  \frac{d}{d\ln{\mu}} S (\mu,\nu) &= \gamma^{s} (\mu,\nu)  S (\mu,\nu)\;,
  & \frac{d}{d\ln{\nu}} S (\mu,\nu) &= \gamma^{s}_{\nu} (\mu,\nu) \otimes S (\mu,\nu) \;.
\end{align}
Therefore we find for the one-loop corresponding impact parameter space quantities
\begin{equation}
  S (\mu,\nu) = 1 + \frac{ \alpha_s(\mu) C_i}{\pi} \lbc
  4 \ln\lp \frac{\mu_E} {\mu}  \rp \ln \lp \frac{\nu}{\mu} \rp - 2 \ln^2\lp \frac{\mu_E} {\mu}  \rp -\frac{\pi^{2}}{12}  
  \rbc  + \mathcal{O}(\alpha_s^2)  \;,
\end{equation}
with
\begin{align}
  \gamma^s (\mu,\nu) &= - 4 \frac{\alpha_s (\mu) C_i}{\pi} \ln \lp \frac{\nu} {\mu} \rp  + \mathcal{O}(\alpha_s^2)   \;, & 
  \gamma^{s}_{\nu} (\mu,\nu)&= 4 \frac{\alpha_s (\mu) C_i}{\pi} \ln \lp \frac{\mu_E} {\mu}  \rp   + \mathcal{O}(\alpha_s^2)\;.
\end{align}

where $\mu_E^{-1} = be^{\gamma_E}$ with $b =|\vec{b}|$ is the impact parameter variable conjugate to $\vec{q}_T$.

\subsection{Resummation}
The vacuum cross section can be resummed by solving the renormalization group equations in $\mu$ and $\nu$ using standard techniques. We will refer the reader to \cite{Gutierrez-Reyes:2019msa} for details on the resummation. In this paper, we are working in a different hierarchy so the details, if not the procedure for resummation, will differ but we postpone presenting the explicit equations till we do a phenomenological study in a future publication.

\section{Medium jet function}
\label{App:JetM}

In this section we present the one loop results for the modified jet function defined in Eq. \ref{JetMed} as the difference of a real (Eq.\ref{JetR}) and virtual (Eq.\ref{JetV}) medium jet function. 
\bea
 \mathcal{J}_{n,R}^{AB}(e_n, \vec{q}_T, \vec{k}_{\perp})&=&\frac{1}{\mathcal{J}^{\perp}_n(e_n,b)k_{\perp}^2}\frac{1}{[p_{\tilde X}^-]^2}\int \frac{d^2q_ne^{-i\vec{q}_n\cdot \vec{b}}}{(2\pi)^2}\langle p_{\tilde X}-k, \bar{X}_n|\delta^2(\mathcal{P}_{\perp})\delta(Q-p^-)\bar\chi_n(0)\frac{\slashed{\bar n}}{2}|0\rangle \nn\\
	&\times&  \langle 0|\chi_n(0)|p_{\tilde X}-k, \bar{X}_n\rangle  \langle \tilde X_n|O_n^A(0)|p_{\tilde X}-k\rangle \langle p_{\tilde X}-k|O_n^B(0)|\tilde X_n\rangle \mathcal{M}_{X_n}
\eea

\bea
\label{Jetv}
\mathcal{J}^{AB}_{n,V}(b) &= & \frac{1}{k_{\perp}^2 \mathcal{J}^{\perp}_n(e_n,b)} \frac{1}{[p_{\tilde Y}^-]^2}\int \frac{d^2q_ne^{-i\vec{q}_n\cdot \vec{b}}}{(2\pi)^2} \langle p_{\tilde Y}+k|O_n^A(0)|\tilde Y_n \rangle \langle \tilde Y_n|O_n^B(0)|p_{\tilde Y}+k \rangle \nn\\
&&\langle p_{\tilde Y}+k,\bar {X}_{n}|\chi_n(0)\frac{\slashed{\bar n}}{2}|0\rangle \langle 0|\bar \chi_n(0)\delta^2(\mathcal{P}_{\perp})\delta^2(Q-\mathcal{P}^-)\mathcal{M}_V|p_{\tilde Y}+k,\bar{X}_n\rangle
\eea
and the medium jet function 
\bea
\mathcal{J}^{AB}_{n,\text{Med}}= \mathcal{J}_{n,R}^{AB}-\mathcal{J}_{n,V}^{AB}
\eea
$\mathcal{J}^{\perp}_n(e_n,b)$ being the vacuum jet function.

The complete one loop corrections include corrections to the Glauber vertex and the corrections to the hard vertex(SCET current). 
In this paper, we will only consider one loop corrections for the hard vertex, while those for the Glauber vertex will be presented in another paper \cite{VarB}. Both of these corrections are equally important and will be needed for any phenomenological analysis.

Therefore, we will evaluate the Glauber collinear current to tree level. The Glauber collinear currents ($O_n^A$, $O_n^{B}$) can be either quark or gluon currents, so we need to consider both cases. However, there will be no contributions proportional to t from the interference term between these currents so we can write the one loop corrections as a sum over the quark and gluon terms.

\begin{itemize} 
\item{Quark Operator}\\
We begin with the quark operator $O_n^A = \bar \chi \frac{\slashed{\bar n}}{2}T^A \chi$.\\
For the quark operator, this will create and annihilate a single quark at tree level so that $|\tilde X_n\rangle \equiv |X_1\rangle $ and $\bar X_n \equiv |X_i\rangle $, with $i \in \{2,\infty\}$.

We can separately consider the contribution from the Real and virtual medium jet functions and combine them later.
\bea
\mathcal{J}_{n,R}^{AB}(e_n,k)&=& 
\frac{1}{\mathcal{J}^{\perp}_n(e_n,b)k_{\perp}^2}\int \frac{d^2q_ne^{-i\vec{q}_n \cdot \vec{b}}}{(2\pi)^2}\Theta(e_n- \frac{8}{Q^2}\sum_{i,j\in \text{gr}}p_{Xi}\cdot p_{Xj})\delta^2(\vec{q}_n- \sum_{i\not\in \text{gr}}\vec{p}^{\perp}_{Xi}-\vec{k}_{\perp})\nn\\
	&&\text{Tr}\Big[T^A T^B\langle p_{X_1}-k,p_{Xi}|\delta^2(\mathcal{P}_{\perp})\delta(Q-p^-)\bar\chi_n(0)\frac{\slashed{\bar n}}{2}|0\rangle \langle 0|
\chi_n(0)| p_{X_1}-k,p_{Xi}\rangle\Big]\nn\\
	\label{Jet}
\eea
we can make a change of variables taking $p_{X_1}-k \rightarrow p_{X_1}$ . We can rearrange the measurement function for the jet mass as 
\bea
 M=\sum_{i,j}p_{Xi}\cdot p_{Xj}&=& \sum_{j\neq 1} p_{X1} \cdot p_{Xj}+  \sum_{i,j\neq 1} p_{Xi} \cdot p_{Xj} \rightarrow  \sum_{i,j}p_{Xi}\cdot p_{Xj}+ \sum_{j} k \cdot p_{Xj} -  k \cdot p_{X_1}\nn
\eea
Now using the fact that $ \sum_{j} p^-_{Xj} =Q$, $ \sum_{j} \vec{p}^{\perp}_{Xj} =0$ and using the power counting of our Glauber mode, 
\bea
M =  \sum_{i,j}p_{Xi}\cdot p_{Xj}+ \frac{1}{2}k^+Q - \frac{1}{2}k^+p_{X_1}^-+\vec{k}_{\perp} \cdot p^{\perp}_{X_1}
\eea

Given the on-shell condition on $p_{X_1}-k$ from Eq. \ref{Jet},
\bea
 k^+p_X^- = 2\vec{p}_{X_1}^{\perp} \cdot \vec{k}_{\perp} +\vec{k}_{\perp}^2
\eea
\bea
M &=& \sum_{i,j}p_{Xi}\cdot p_{Xj} +\frac{Q}{2p_{X_1}^-}2\vec{p}_{X_1}^{\perp} \cdot \vec{k}_{\perp} +\frac{\vec{k}_{\perp}^2}{2p_{X_1}^-}\left(Q-p_{X_1}^-\right)
\eea
so that our jet function becomes
\small
\bea
 && \mathcal{J}^{AB}(e_n,k)=\frac{1}{\mathcal{J}^{\perp}_n(e_n,b)k_{\perp}^2}\int \frac{d^2q_ne^{-i\vec{q}_n \cdot \vec{b}}}{(2\pi)^2}\Theta\Big(e_n - \Big[8\frac{\vec{k}_{\perp}\cdot p_{X_1}^{\perp}}{Qp_{X_1}^-}+8\frac{k_{\perp}^2}{2p_X^-Q^2}(Q-p_{X_1}^-)\Big]_{X_1\in \text{gr}}\nn\\
&-&\frac{8}{Q^2}\sum_{i,j \in \text{gr}}p_{Xi}\cdot p_{Xj}\Big)\delta^2(\vec{q}_n- \sum_{i\not\in \text{gr}}\vec{p}_{Xi,\perp}-\vec{k}_{\perp}\Big|_{X_1 \in \text{gr}})\langle X_i|\delta^2(\mathcal{P}_{\perp})\delta(Q-p^-)\bar\chi_n(0)\frac{\slashed{\bar n}}{2}|0\rangle \langle 0|\chi_n(0)|X_i\rangle\nn\\
	\label{JetN}
\eea
\normalsize
So the definition of the jet mass measurement is altered from its standard definition by a function of $k_{\perp}$ due to the interaction of the parton $X_1$ with the medium. Even though we have worked this out for the case of jet mass measurement, this is a general property for all jet substructure observables. $p_{X_1}$ here is the momentum of a quark so that when we compute the jet function, it will automatically sum over all the quarks in the jet as it should. When we include the gluon-medium interaction, we will have a similar sum over gluon states. 
\bea
\mathcal{J}_{n,R}^{AB,(0)}(e_n,\vec{k}_{\perp},\vec{b}) = \frac{\frac{1}{2}\delta^{AB}}{k_{\perp}^2}e^{-i \vec{k}_{\perp} \cdot \vec{b}}  
\eea
The vacuum jet function which appears in the denominator is just 1 at tree level.

At one loop, I have both real and virtual diagrams. The gluon emitted as a radiative correction interacts with the medium and acquires a mass $m_D$ with the hierarchy $m_D  \ll e_n$ since we are working in a weak coupling regime. If the jet function is IR finite, then the $m_D$ scale is irrelevant and we can use dimensional regularization which does not distinguish between UV and IR divergences. However, we expect that the medium will induce non-trivial Infra-Red physics of this function, it is no longer guaranteed that this function is IR finite, in the sense that it can be sensitive to scale $m_D$. We are dividing out by the vacuum jet function which, at one loop is equivalent to subtracting the one loop result in impact parameter(b) space. This immediately tells us that the virtual diagrams simply cancel out between the numerator and denominator and the answer will just be the difference between the real diagrams which we now consider.
There are two Feynman diagram that contribute  as shown in Fig.\ref{Real}.
\bea
\mathcal{J}_{n,R}^{AB,(1)} =\frac{\frac{1}{2}\delta^{AB}}{k_{\perp}^2}\Bigg\{(R_a(\vec{b}) - e^{-i\vec{k}_{\perp} \cdot \vec{b}}R_a^{\text{Vac}}(\vec{b}))+(R_b(\vec{b})-e^{-i\vec{k}_{\perp} \cdot \vec{b}}R_b^{\text{Vac}}(\vec{b}))\Bigg\}\equiv \frac{\frac{1}{2}\delta^{AB}}{k_{\perp}^2}R(b)\nn
\eea
\begin{figure}
     \centering
     \subfloat[][]{\includegraphics[width=.6\linewidth]{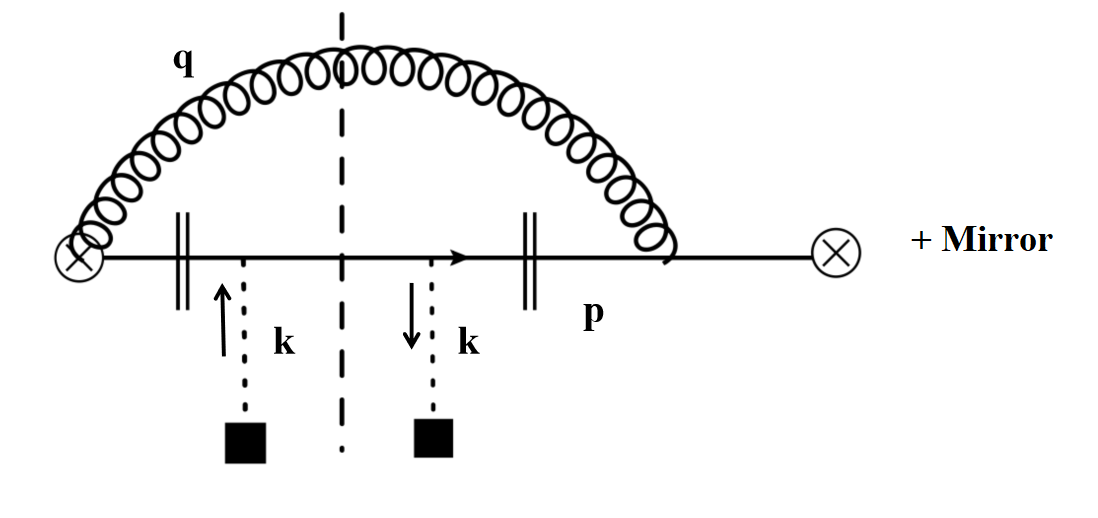}\label{a}}
     \subfloat[][]{\includegraphics[width=.4\linewidth]{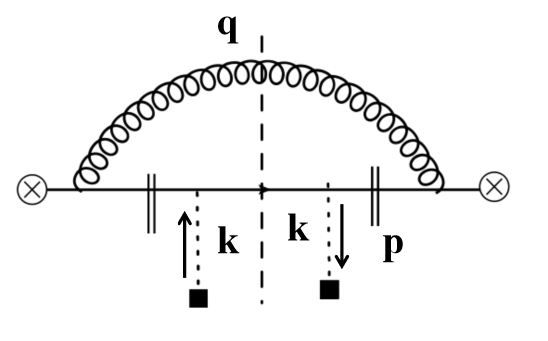}\label{b}}
     \caption{Real emission diagrams}
     \label{Real}
\end{figure}

Depending on which partons pass the grooming condition, we have three contributions just as for the vacuum diagram. For the diagram (a), we have
\small
\bea
 R_a &=& 8g^2C_F \int \frac{d^2q_{n}}{(2\pi)^2}e^{-i\vec{q}_n\cdot \vec{b}}\int d^4p \delta^+(p^2)\int \frac{d^4q}{(2\pi)^{3}}\delta^+(q^2-m_D^2)\delta(Q-p^--q^-)\delta^2(\vec{q}_{\perp}+\vec{p}_{\perp})\frac{p^-(p^-+q^-)}{q^-(p+q)^2}\nn\\
&&\Big\{\Theta\left(e_n-8\frac{\vec{k}_{\perp}\cdot \vec{p}_{\perp}}{Qp^-}-8\frac{k_{\perp}^2(Q-p^-)}{2p^-Q^2}-\frac{8}{2Q}\left(p^++q^+\right)\right)\delta^2(\vec{q}_{n}-\vec{k}_{\perp})\Theta(q^- -Qz_c)\Theta(p^--Qz_c)\nn\\
&+& \Theta(Qz_{c}-q^-)\Theta(p^--Qz_{c})\delta^2(\vec{q}_n-\vec{q}_{\perp}-\vec{k}_{\perp})+\Theta(Qz_{c}-p^-)\Theta(q^--Qz_{c})\delta^2(\vec{q}_n-\vec{p}_{\perp})\Big\}
\eea
\normalsize
we can shift $\vec{q}_n \rightarrow \vec{q}_n +\vec{k}_{\perp}$ and subtract the corresponding vacuum diagram in Fig. \ref{vac}. Doing the same for diagram (b), we get
\small
\bea
 R&=& 4g^2C_F\int \frac{d^2q_{n}}{(2\pi)^2}e^{-i(\vec{q}_n+\vec{k}_{\perp})\cdot \vec{b}}\int \frac{d^4q}{(2\pi)^{3}}\delta^+(q^2-m_D^2)\Big\{\frac{(Q-q^-)}{q^-\left(q^++ \frac{q_{\perp}^2}{Q-q^-}\right)}+\frac{q_{\perp}^2}{2(Q-q^-)^2(q^++\frac{q_{\perp}^2}{Q-q^-})^2}\Big\}\nn\\
&&\Big\{\left(\Theta\left(e_n-8\frac{\vec{k}_{\perp}\cdot \vec{p}_{\perp}}{Qp^-}-8\frac{k_{\perp}^2(Q-p^-)}{2p^-Q^2}-\frac{8}{2Q}\left( \frac{q_{\perp}^2}{Q-q^-}+q^+\right)\right)-\Theta(e_n-\frac{8}{2Q}\left( \frac{q_{\perp}^2}{Q-q^-}+q^+\right))\right) \nn\\
&&\delta^2(\vec{q}_{n})\Theta(q^- -Qz_c)\Theta(p^--Qz_c)+\Theta(Qz_{c}-p^-)\Theta(q^--Qz_{c})\Big[\delta^2(\vec{q}_n+\vec{k}_{\perp}+\vec{q}_{\perp})-\delta^2(\vec{q}_n+\vec{q}_{\perp})\Big]\Big\}\nn
\eea
\normalsize
The other piece of the medium jet function is $\mathcal{J}_{n,V}^{AB}$ defined in Eq. \ref{Jetv}. Once again we consider the quark operator and evaluate the Glauber current at tree level, which leads to 
\bea
\mathcal{J}^{AB}_{n,V}(b) &= & \frac{\frac{1}{2}\delta^{AB}}{k_{\perp}^2 \mathcal{J}^{\perp}_n(e_n,b)}\int \frac{d^2q_ne^{-i\vec{q}_n\cdot \vec{b}}}{(2\pi)^2} \langle p_{\tilde Y}+k,\bar {X}_{n}|\chi_n(0)\frac{\slashed{\bar n}}{2}|0\rangle \nn\\
&&\langle 0|\bar \chi_n(0)\delta^2(\mathcal{P}^{\perp})\delta^2(Q-\mathcal{P}^-)\mathcal{M}_V|p_{\tilde Y}+k,\bar{X}_n\rangle \nn\\
&=&  \frac{\frac{1}{2}\delta^{AB}}{k_{\perp}^2}
\eea
where the numerator is simply proportional to the the vacuum jet function and hence cancels out with the denominator leaving behind a very simple result. 
Therefore, for the Quark Glauber operator, our result is 
\bea
\mathcal{J}_{n,\text{Med}}^{AB,q}&=& \frac{\frac{1}{2}\delta^{AB}}{k_{\perp}^2}\Big\{e^{-i \vec{k}_{\perp} \cdot \vec{b}}  -1 +R(b)\Big\} 
\eea

R(b) evaluates to 
\small
\bea
 &R&= \frac{\alpha_s C_F e^{-i\vec{k}_{\perp}\cdot \vec{b}}}{2 \pi} \Bigg[\int_{z_c}^{1-z_c}dz p_{gq}(z)\Big\{\Theta(z_c-\frac{y}{e_n+y})\ln \left(\frac{-B(z)}{e_n(1-z)}\right)-\Theta(z_c-\frac{e_n}{e_n+y})\ln \frac{B(z)e_n}{yM}\Big\}\nn\\
&+&\Theta( \frac{e_n}{e_n+y}-z_c)\Theta(\frac{y}{e_n+y}-z_c)\Big\{\int_{z_c}^{\frac{e_n}{e_n+y}}dz p_{gq}(z)\ln \left(\frac{-B(z)}{e_n(1-z)}\right)-\int_{\frac{e_n}{e_n+y}}^{1-z_c}dz p_{gq}(z)\ln \frac{B(z)e_n}{yM}\Big\}\Bigg]\nn\\
&+& \frac{\alpha_sC_F}{2\pi}\left(e^{-i\vec{k}_{\perp}\cdot \vec{b}}-1\right)\int_{1-z_c}^{1}p_{gq}(z)\ln \frac{m_D^2b^2e^{2\gamma_E}(1-z)}{4}
\eea
\normalsize
where 
\bea
p_{gq}(z) = \frac{1+ (1-z)^2}{z}, \ \ \ B(z)=zy- e_n(1-z) , \ \ \  y = \frac{4k_{\perp}^2}{Q^2}, \ \ \ M =\frac{4m_D^2}{Q^2} 
\eea 
The integral over z can be done analytically with little difficulty. However, the resulting final expression is long and not very illuminating, so we present the result as an integral over z. The result is finite and there are no UV or rapidity divergences. However, it is sensitive to the IR cut-off $m_D$.

\item{Gluon Operator}

We will now consider the contribution to the one loop modified jet function from the gluon operator $O_{n,g}^A= \frac{i}{2}f^{ABC}\mathcal{B}_{n\perp,\mu}^B\frac{\bar n}{2}\cdot(\mathcal{P}+\mathcal{P}^{\dagger})\mathcal{B}_{n\perp}^{C\mu}$, which creates and annihilates a single gluon at tree level.

Following the same series of steps as for the quark operator, we arrive at 
\small
\bea 
&&\mathcal{J}_{n,R}^{AB,g}(e_n,\vec{k}_{\perp}) = \frac{1/2}{k_{\perp}^2\mathcal{J}^{\perp}_n(e_n,b)}\int \frac{d^2q_ne^{-i\vec{q}_n\cdot \vec{b}}}{(2\pi)^2}\Theta(e_n - \Big[8\frac{\vec{k}_{\perp}\cdot p_{X_1}^{\perp}}{Qp_{X_1}^-}+8\frac{k_{\perp}^2}{2p_{X_1}^-Q^2}(Q-p_{X_1}^-)\Big]_{X_1\in \text{gr}}\nn\\
&-&\frac{8}{Q^2}\sum_{i,j \in \text{gr}}p_{Xi}\cdot p_{Xj})\delta^2(\vec{q}_n- \sum_{i\not\in \text{gr}}\vec{p}_{Xi}^{\perp}-\vec{k}_{\perp}\Big|_{X_1 \in \text{gr}})\langle p_{Xi}|\delta^2(\mathcal{P}^{\perp})\delta(Q-p^-)\bar\chi_n(0)\frac{\slashed{\bar n}}{2}|0\rangle \langle 0|\chi_n(0)|p_{Xi}\rangle \nn\\
	&\times& \frac{f^{Abc}f^{Bbc}\epsilon^{\mu}(p_{X_1})\epsilon^{*\nu}(p_{X_1})}{(p_{X_1}^-)^2}\Big[(p_{X_1}^-)^2g_{\perp}^{\mu \nu}-\bar{n}^{\mu}p_{X_1}^{\perp {\nu}}p_{X_1}^--\bar{n}^{\mu}\bar{n}^{\nu}p^2_{X_1,\perp}-\bar{n}^{\nu}p_{X_1}^-p_{X_1\perp}^{\mu}\Big]
\eea
\normalsize
where $p_{X_1}$ is now the momentum of a gluon.  The contribution from the gluon operator only starts at O($\alpha_s$), hence for a one loop calculation, we can evaluate the vacuum $J_n(e_n,b)$ in the denominator to tree level.
\bea
 \mathcal{J}_{n,R}^{AB,g(1)}(e_n,\vec{k}_{\perp},b) &=& \frac{\frac{1}{2}\delta^{AB}}{k_{\perp}^2}R^g
\eea
We can now compute the one loop results which correspond to the diagram Fig. \ref{RGC}. However, diagram (a) reduces to 0 and we only get a contribution form (b). This gives us 
\small
\bea
\label{Rg}
R_g&=& 4g^2N_c \int \frac{d^2q_{n}}{(2\pi)^2}e^{-i\vec{q}_n\cdot \vec{b}}\int d^4p \delta^+(p^2)\int \frac{d^4q}{(2\pi)^{3}}\nn\\
&&\delta^+(q^2-m_D^2)\delta(Q-p^--q^-)\delta^2(\vec{q}_{\perp}+\vec{p}_{\perp})\frac{Q^2+(p^-)^2}{p^-(q^-)^2[(p+q)^2]^2}\nn\\
&&\Big\{\Theta\left(e_n-8\frac{\vec{k}_{\perp}\cdot \vec{q}_{\perp}}{Qq^-}-8\frac{k_{\perp}^2(Q-q^-)}{2q^-Q^2}-\frac{8}{2Q}\left(p^++q^+\right)\right)\delta^2(\vec{q}_{n}-\vec{k}_{\perp})\Theta(q^- -Qz_c)\Theta(p^--Qz_c)\nn\\
&+& \Theta(Qz_{c}-q^-)\Theta(p^--Qz_{c})\delta^2(\vec{q}_n-\vec{q}_{\perp})+\Theta(Qz_{c}-p^-)\Theta(q^--Qz_{c})\delta^2(\vec{q}_n-\vec{p}_{\perp}-\vec{k}_{\perp})\Big\}
\eea
\normalsize
\begin{figure}
  \includegraphics[width=\textwidth]{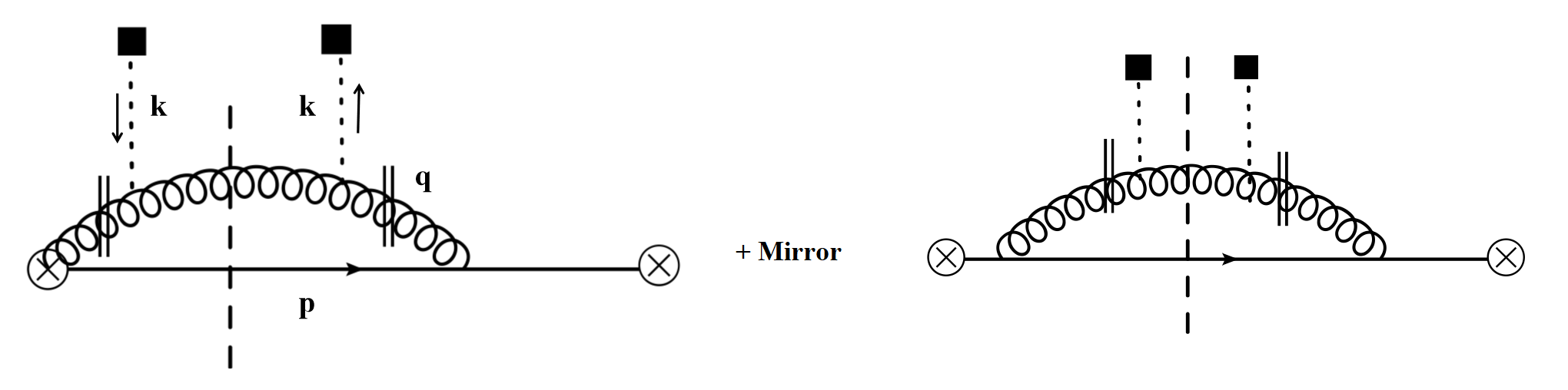}
  \caption{Gluon real interaction with the medium}
  \label{RGC}
\end{figure}

We can now do a similar analysis for the gluon operator contribution to $\mathcal{J}_{n,V}^{AB}$ which at one loop yields
\bea
 \mathcal{J}_{n,V}^{AB,g(1)}(e_n,\vec{k}_{\perp},b) &=& \frac{\frac{1}{2}\delta^{AB}}{k_{\perp}^2}V^g
\eea
These are the diagrams where the Glauber insertion happens on the same side of the cut as shown in Fig. \ref{VMGa}. Given that the gluon is on-shell, this will merely give us the real vacuum diagram dressed with the virtual interaction of the gluon with the medium.
\begin{figure}
  \includegraphics[width=\textwidth]{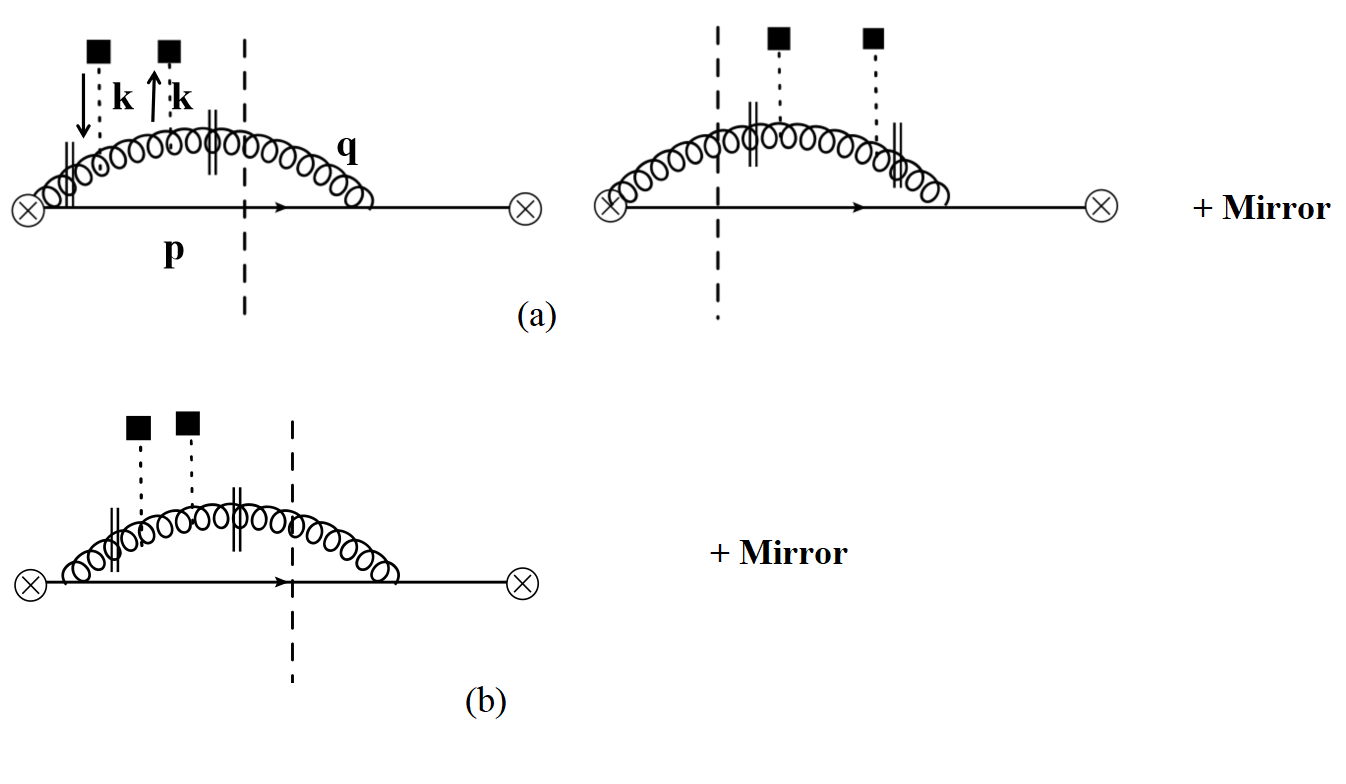}
  \caption{Gluon virtual interaction with the medium}
  \label{VMGa}
\end{figure}

\small
\bea
	V_{g}&=& 4g^2N_c \int \frac{d^2q_{n}}{(2\pi)^2}e^{-i\vec{q}_n\cdot \vec{b}}\int d^4p \delta^+(p^2)\int \frac{d^4q}{(2\pi)^{3}}\delta^+(q^2-m_g^2)\delta(Q-p^--q^-)\delta^2(\vec{q}_{\perp}+\vec{p}_{\perp})\nn\\
&&\frac{Q^2+(p^-)^2}{p^-(q^-)^2[(p+q)^2]^2}\Big\{\Theta\left(e_n-\frac{8}{2Q}\left(p^++q^+\right)\right)\delta^2(\vec{q}_{n})\Theta(q^- -Qz_c)\Theta(p^--Qz_c)\nn\\
&+& \Theta(Qz_{cut}-q^-)\Theta(p^--Qz_{cut})\delta^2(\vec{q}_n-\vec{q}_{\perp})+\Theta(Qz_{cut}-p^-)\Theta(q^--Qz_{cut})\delta^2(\vec{q}_n-\vec{p}_{\perp})\Big\}\nn\\
\eea
\normalsize
which is just the same as $R_g$ ( Eq .\ref{Rg}) in the limit $\vec{k}_{\perp} \rightarrow 0$. We can then combine the two terms to give us the full contribution from the gluon current 
\bea
 \mathcal{J}_{n,\text{Med}}^{AB,g}(e_n,\vec{k}_{\perp},b) &=& \frac{\frac{1}{2}\delta^{AB}}{k_{\perp}^2}\Big\{R_g-V_g\Big\}
\eea
where $R_g$ evaluates to 
\small
\bea
 R_g&=& \frac{\alpha_s N_c e^{-i\vec{k}_{\perp}\cdot \vec{b}}}{2 \pi} \Bigg[\int_{z_c}^{1-z_c}dz p_{gq}(z)\Big\{\Theta(z_c-\frac{y}{e_n+y})\ln\frac{A(z)}{M}-\Theta(z_c-\frac{e_n}{e_n+y})\ln \left(\frac{-A(z)}{y(1-z)}\right)\Big\}\nn\\
&-&\Theta( \frac{e_n}{e_n+y}-z_c)\Theta(\frac{y}{e_n+y}-z_c)\Big\{\int_{z_c}^{\frac{y}{e_n+y}}dz p_{gq}(z)\ln \left(\frac{-A(z)}{y(1-z)}\right)-\int_{\frac{y}{e_n+y}}^{1-z_c}dz p_{gq}(z)\ln \frac{A(z)}{M}\Big\}\Bigg]\nn\\
&-& \frac{\alpha_sN_c}{2\pi}e^{-i\vec{k}_{\perp}\cdot \vec{b}}\int_{1-z_c}^{1}p_{gq}(z)\ln \frac{m_D^2b^2e^{2\gamma_E}(1-z)}{4}
\eea 
\normalsize
which again is sensitive to $\ln m_D$ and 
\bea
A(z) =e_nz-(1-z)y 
\eea
$V_g$ is obtained by simply evaluating $R_g$ at $k_{\perp}=0$.
\end{itemize}


\end{document}